\documentclass[aps,twocolumn,preprintnumbers,floats]{revtex4}\def\@cite#1#2{\textsuperscript{[{#1\if@tempswa , #2\fi}]}}
\usepackage{graphicx,color,dcolumn,booktabs,bm}
\usepackage{amsmath}
\usepackage{graphicx}
\usepackage{float}
\usepackage{longtable,lscape}
\usepackage{txfonts}
\usepackage{mathrsfs}
\usepackage{overpic}
\usepackage{epsfig}
\usepackage{amssymb}
\usepackage{bbding}
\usepackage{rotating}
\usepackage{epstopdf}
\usepackage{appendix}
\usepackage{indentfirst}
\usepackage{feynmf}   
\usepackage{slashed}  
\usepackage{cases}
\usepackage{color}
\usepackage{booktabs,rotating}
\usepackage{ulem}
\usepackage{multirow}
\usepackage{graphicx,color,dcolumn,booktabs,bm}
\usepackage{cases}
\usepackage{array}
\usepackage{booktabs} 
\usepackage{amsmath}  
\usepackage{enumerate}
\usepackage{braket}

\graphicspath{{Figures/}} %

\usepackage[colorlinks, citecolor=blue,anchorcolor=red,menucolor=red, linkcolor=red,filecolor=red,runcolor=red,urlcolor=blue,frenchlinks=red]{hyperref}
\usepackage{CJK}

\newcommand{\vlab}{\mbox{\boldmath$\lambda$\unboldmath}}
\newcommand{\vsig}{\mbox{\boldmath$\sigma$\unboldmath}}
\newcommand{\vxi}{\mbox{\boldmath$\xi$\unboldmath}}

\begin{document}

\title{All-heavy tetraquarks with different flavors}

\author{Wei-Xiang Wang$^{1}$, Lin-Qin Xie$^{1}$, Jun-Jie Liu$^{1}$, Zhi-Biao Liang$^{1}$,
Ming-Sheng Liu$^{2}$~\footnote{E-mail: liumingsheng@email.tjut.edu.cn},
Xian-Hui Zhong$^{1,3}$~\footnote{E-mail: zhongxh@hunnu.edu.cn}}
\affiliation{ 1) Department of Physics, Hunan Normal University, and Key Laboratory of Low-Dimensional Quantum Structures and Quantum Control of Ministry of Education, Changsha 410081, China }
\affiliation{ 2) Tianjin Key Laboratory of Quantum Optics and Intelligent Photonics, School of Science, Tianjin University of Technology, Tianjin 300384, China}
\affiliation{ 3) Synergetic Innovation Center for Quantum Effects and Applications (SICQEA), Hunan Normal University, Changsha 410081, China}

\begin{abstract}
In a nonrelativistic potential quark model framework, we carry out a precise calculation of the mass spectrum
of the all-heavy tetraquarks with different flavors, $bb\bar{b}\bar{c}$, $cc\bar{c}\bar{b}$, $bb\bar{c}\bar{c}$, and $bc\bar{b}\bar{c}$, by adopting the explicitly correlated Gaussian method. A complete mass spectrum for the $1S$ states is obtained. For the $bb\bar{b}\bar{c}$, $cc\bar{c}\bar{b}$, $bb\bar{c}\bar{c}$, and $bc\bar{b}\bar{c}$ systems, the $1S$ states are predicted to lie in the mass ranges of $ \sim(16.06,16.14)$, $\sim(9.65,9.74)$, $\sim(12.89,12.94)$, and $\sim(12.75,12.99)$~GeV, respectively.
Moreover, by using the obtained masses and wave functions, we evaluate the fall-apart decay properties within a quark-exchange model.
The results show that the $1S$ states of the all-heavy tetraquarks with different flavors may have narrow fall-apart decay widths,
which range from a few tenths to several MeV. Some all-heavy tetraquarks with different flavors could be search for
at LHC in their optimal fall-apart decay channels, such as $\Upsilon J/\psi$, $\Upsilon B_c^-$, and $J/\psi B_c^+$.
\end{abstract}

\maketitle

\section{introduction}

Among exotic hadrons, the all-heavy tetraquarks have attracted considerable attention as a system of significant interest. Since light mesons cannot be exchanged, all-heavy tetraquarks are considered ideal systems for exploring genuine compact tetraquark states.
In 2020, the LHCb collaboration observed a narrow structure $X(6900)$ in the di-$J/\psi$ invariant mass spectrum~\cite{LHCb:2020bwg}.
Its existence was later confirmed independently by the CMS~\cite{CMS:2023owd} and ATLAS~\cite{ATLAS:2023bft} collaborations.
Furthermore, the CMS also observed two additional new structures $X(6600)$ and $X(7100)$ in the di-$J/\psi$ invariant mass spectrum~\cite{CMS:2023owd}. These structures could be interpreted as tetraquark states with four charm quarks,
$cc\bar{c}\bar{c}$~\cite{2Bedolla:2019zwg,Wu:2016vtq,1Wang:2019rdo,ms100:2019,Iwasaki:1975pv,Chao:1980dv,Debastiani:2017msn,Chen:2016jxd,Mutuk:2021hmi}.
The CMS and LHC collaborations have also been dedicated to searching for the fully bottomed tetraquarks $bb\bar{b}\bar{b}$,
however, no significant signals have been found so far~\cite{CMS:2016liw,CMS:2020qwa,LHCb:2018uwm}.

Besides $cc\bar{c}\bar{c}$ and $bb\bar{b}\bar{b}$,
there also exist other all-heavy tetraquarks containing both charm and bottom quarks,
$bb\bar{b}\bar{c}$/$bc\bar{b}\bar{b}$, $cc\bar{c}\bar{b}$/$bc\bar{c}\bar{c}$, $bb\bar{c}\bar{c}$/$cc\bar{b}\bar{b}$, and $bc\bar{b}\bar{c}$. The new discovery of several $cc\bar{c}\bar{c}$ candidates at LHC indicates the experimental investigation of the other all-heavy tetraquark states
containing both charm and bottom quarks also exhibits considerable potentials.
In fact, the LHC has also demonstrated powerful capabilities in searching for hadrons containing both
charm and bottom quarks. For example, several excited $B_c$ states~\cite{CMS:2019uhm,LHCb:2019bem,LHCb:2025uce},
and evidence of the doubly heavy baryon $\Xi_{bc}$~\cite{LHCb:2022fbu} were observed at LHC, recently.
The experimental progress stimulated theoretical research interest
in these all-heavy tetraquarks with different flavors. In recent several years, numerous studies of
the mass spectra have been carried out within many models and approaches, such as,
various nonrelativistic constituent quark models~\cite{ms100:2019,4Gordillo:2020sgc,1Wang:2019rdo,3Deng:2020iqw,11Hu:2022zdh,12Zhang:2022qtp,21Wu:2024hrv,27Ortega:2025lmo,35An:2022qpt},
diquark models~\cite{5Faustov:2020qfm,2Bedolla:2019zwg,13Faustov:2022mvs,15Galkin:2023wox,Mohan:2026blk,Mutuk:2022nkw}, QCD sum rules~\cite{7Yang:2021zrc,9Wang:2021taf,14Chen:2022mcr,34Agaev:2025qgg,33Agaev:2025wyf,31Agaev:2025nkw,32Agaev:2025did,29Agaev:2025fwm,28Agaev:2025wdj,
25Agaev:2024uza,24Agaev:2024qbh,23Agaev:2024mng,22Agaev:2024wvp,Agaev:2023tzi,20Agaev:2024xdc,19Agaev:2024pil,18Agaev:2024pej},
color-magnetic models~\cite{Wu:2016vtq,3Deng:2020iqw,6Weng:2020jao,10Zhuang:2021pci}, Bethe-Salpeter equation method~\cite{30Wang:2025apq},
the flux-tube model~\cite{3Deng:2020iqw}, bosonic algebraic approach~\cite{8Majarshin:2021hex}, conditional generative adversarial network (CGAN)~\cite{26Malekhosseini:2025hyx}, pNRQCD method~\cite{16Assi:2023dlu}, heavy meson exchanged model~\cite{17Liu:2023gla}, and so on. However, comparing existing model calculations, one can find that there is a strong model dependency in the results.

For the all-heavy tetraquark systems with different flavors, $bb\bar{b}\bar{c}$/$bc\bar{b}\bar{b}$, $cc\bar{c}\bar{b}$/$bc\bar{c}\bar{c}$, $bb\bar{c}\bar{c}$/$cc\bar{b}\bar{b}$, the $1S$-wave mass spectra were preliminarily studied within a nonrelativistic quark potential model by our group in 2019~\cite{ms100:2019}. In the calculations,
the oscillator parameter of the trial wave function was approximately treated as a quark mass independent parameter when solving the mass spectrum via the variational method. However, such a treatment should result in a serious incompleteness of the trial wave function
for the all-heavy tetraquark systems with different quark flavors.
In the present work, to improve the completeness of the trial wave function, and obtain more reliable predictions of the mass spectra, we revise the $bb\bar{b}\bar{c}$/$bc\bar{b}\bar{b}$, $cc\bar{c}\bar{b}$/$bc\bar{c}\bar{c}$, $bb\bar{c}\bar{c}$/$cc\bar{b}\bar{b}$ systems by adopting the correlated Gaussian functions~\cite{Varga:1995dm,Varga:1997xga,Mitroy:2013eom} as the radial wave function basis.
This method is known to be effective and accurate for solving few-body problems.

Considering the fact that the obtained $1S$-wave all-heavy tetraquark states with different flavors
lie far above the dissociation into two ground meson threshold, we further evaluate
their fall-apart decay properties within a quark-exchange model~\cite{Barnes:1991em,Barnes:2000hu}.
The present study on the fall-apart decay properties of the
$bb\bar{b}\bar{c}$, $cc\bar{c}\bar{b}$, $bb\bar{c}\bar{c}$, and $bc\bar{b}\bar{c}$ systems
is a continuation of our previous work on the $cc\bar{c}\bar{c}$ and $bb\bar{b}\bar{b}$ systems~\cite{liu:2020eha}.
By the study of their decay properties,
we expect to provide useful decay channels for future experimental observing.
The study of decay properties of all-heavy tetraquark states with different flavors
is relatively scarce. Only a few research groups have carried out
some exploration on this matter with different methods,
such as the complex scaling method~\cite{21Wu:2024hrv}, real scaling method~\cite{11Hu:2022zdh},
the coupled-channels method~\cite{27Ortega:2025lmo}, CGAN framework~\cite{26Malekhosseini:2025hyx}, QCD sum rules~\cite{34Agaev:2025qgg,33Agaev:2025wyf,31Agaev:2025nkw,32Agaev:2025did,29Agaev:2025fwm,28Agaev:2025wdj,
25Agaev:2024uza,24Agaev:2024qbh,23Agaev:2024mng,22Agaev:2024wvp,Agaev:2023tzi,20Agaev:2024xdc,19Agaev:2024pil,18Agaev:2024pej}, and so on.
There are strong model dependencies of the decay properties. For example,
the tetraquarks $bb\bar{b}\bar{c}$ and $cc\bar{c}\bar{b}$
are predicted to be broad structures with a width of $\sim100$ MeV within the QCD sum rules ~\cite{25Agaev:2024uza,32Agaev:2025did,33Agaev:2025wyf,34Agaev:2025qgg},
while narrow structures with a width of about several MeV within the real scaling method~\cite{11Hu:2022zdh}.

This paper is organized as follows.
In Sec.~\ref{Framework}, the theoretical framework is briefly introduced.
In Sec.~\ref{results}, the numerical results and discussions of
all-heavy tetraquarks with different flavors are presented.
Finally, a short summary is given in Sec.~\ref{summary}.

%
%
%
%

\section{FRAMEWORK}\label{Framework}

\subsection{Mass spectrum}

\subsubsection{Hamiltonian}

In this work, to describe the tetraquark system we adopt a nonrelativistic Hamiltonian~\cite{ms100:2019}, i.e.
\begin{eqnarray}\label{Hamiltonian}
H &=&
    \sum_{i=1}^{4}\left(m_i +T_i\right)-T_G+\sum_{i<j}V_{ij}(r_{ij}),
\end{eqnarray}
where $m_i$ and $T_i$ stand for the mass and kinetic energy of the $i$-th quark, respectively.
$T_G$ is the center-of-mass kinetic energy.
$V_{ij}(r_{ij})$ represents the effective potentials between the $i$-th and $j$-th quarks with a distance $r_{ij}\equiv|\boldsymbol{r}_i-\boldsymbol{r}_j|$.
In this work, we adopt a widely used potential form for $V_{ij}(r_{ij})$~\cite{Eichten:1978tg,Capstick:1986ter,Godfrey:1985xj}, i,e.,
\begin{eqnarray}\label{OGE}
V_{ij}(r_{ij})&=&
-\frac{3}{16}(\vlab_i\cdot\vlab_j)\left\{br_{ij}-\frac{4}{3}\frac{\alpha_{ij}}{r_{ij}}\right.\nonumber\\
&&\left.+\alpha_{ij}\cdot\frac{\sigma^3_{ij}e^{-\sigma^2_{ij}r^2_{ij}}}{\pi^{1/2}}\cdot\frac{8}{9m_im_j}(\vsig_i\cdot\vsig_j)\right\},
\end{eqnarray}
where the first, second, and last terms correspond to the linear confinement part $V^{\mathrm{Conf}}$, color-Coulomb part $V^{\mathrm{Coul}}$, and chromo-magnetic (spin-spin) part $V^{\mathrm{SS}}$, respectively. $\vlab_i$ and $\vsig_i$ stand for the spin and color operator of the $i$-th quark, respectively.
The $b$ is the slope parameter of the confinement potentials,
while $\alpha_{ij}$ are the strong coupling constants.

The nine parameters $m_{c/b}$, $\alpha_{cc/bb/bc}$, $\sigma_{cc/bb/bc}$, and $b$
have been determined by fitting the $c\bar{c}$, $b\bar{b}$, and ${b\bar{c}}$ spectrum
in our previous works~\cite{Deng:2016stx,ms100:2019,Li:2019tbn}.
The parameter set, along with the theoretical masses compared with the experimental data~\cite{ParticleDataGroup:2024cfk}, are listed in Table~\ref{qrkpr}.


\begin{table}[hptb]
	\caption{The quark model parameters determined by fitting the
meson mass spectra. The unit of the meson masses is MeV. }\label{qrkpr}
	\begin{tabular}{cccccc}
		\hline\hline
\multicolumn{6}{c}{Quark model parameters}\\ \hline
\multicolumn{3}{c}{$m_c/m_b$(GeV)}&\multicolumn{3}{c}{1.483/4.852}\\
\multicolumn{3}{c}{$\alpha_{cc}/\alpha_{bc}/\alpha_{bb}$}&\multicolumn{3}{c}{0.5461/0.5021/0.4311}\\
\multicolumn{3}{c}{$\sigma_{cc}/\sigma_{bc}/\sigma_{bb}$(GeV) }&\multicolumn{3}{c}{1.1384/1.3000/2.3200}\\
\multicolumn{3}{c}{$b$(GeV$^2$)}&\multicolumn{3}{c}{0.1425}\\ \hline\hline
\multicolumn{6}{c}{Mass spectrum}\\
\hline
 Meson      & Theor. & Exp.~\cite{ParticleDataGroup:2024cfk} &         Meson         & Theor. & Exp.~\cite{ParticleDataGroup:2024cfk} \\
\hline
		 $J/\psi(1S)$   & 3097 & 3097  &   $B_{c}(2S)^{\pm}$   & 6871  & 6871  \\
		$\eta_{c}(1S)$  & 2983 & 2984  & $\Upsilon\text{(1S)}$ & 9460  & 9460  \\
		  $\psi(2S)$    & 3679 & 3686  &    $\eta_{b}(1S)$     & 9390  & 9399  \\
		$\eta_{c}(2S)$  & 3635 & 3638  &    $\Upsilon(2S)$     & 10024 & 10023 \\
		$\chi_{c0}(1P)$ & 3415 & 3415  &    $\eta_{b}(2S)$     & 10005 & 9999  \\
		$\chi_{c1}(1P)$ & 3516 & 3511  &      $h_{b}(1P)$      & 9941  & 9899  \\
		$\chi_{c2}(1P)$ & 3552 & 3556  &    $\chi_{b0}(1P)$    & 9859  & 9859  \\
		  $h_{c}(1P)$   & 3522 & 3525  &    $\chi_{b1}(1P)$    & 9933  & 9893  \\
		  $B_{c}^{+}$   & 6271 & 6274  &    $\chi_{b2}(1P)$    & 9957  & 9912  \\ \hline\hline
	\end{tabular}
\end{table}

\begin{table*}[htp]
\begin{center}
\caption{\label{configurations} Configurations of all-heavy tetraquarks with different flavors, where $\{~\}$ and $[~]$ denote the symmetric and antisymmetric flavor wave functions of the two quarks (antiquarks) subsystems, respectively. The subscripts and superscripts are the spin quantum numbers and representations of the color SU(3) group, respectively.}
\renewcommand\arraystretch{1.5}
\tabcolsep=0.45cm
\begin{tabular}{cccccccccccc}\hline\hline
 System              ~~~~~~~& $J^{P(C)}$  ~~~~~~~& \multicolumn{3}{c}{ \underline{~~~~~~~~~~~~~~~~~~~~~~~~~~~~~~~~~~~~~~~~~~~~~~~~~~~~~~~~~~~Configuration~~~~~~~~~~~~~~~~~~~~~~~~~~~~~~~~~~~~~~~~~~~~~~~~~~~~~~~~~~~~~~} }                        \\
\hline
$bb\bar{b}\bar{c}$  ~~~~~~~& $0^{+}$                ~~~~~~~& $|(bb)^6_0\{\bar{b}\bar{c}\}^{\bar{6}}_0\rangle^0_0$      ~& $|(bb)^{\bar{3}}_1\{\bar{b}\bar{c}\}^3_1\rangle^0_0$
                                                                ~&            $\cdot\cdot\cdot$                          \\
                     ~~~~~~~& $1^{+}$                ~~~~~~~& $|(bb)^6_0\{\bar{b}\bar{c}\}^{\bar{6}}_1\rangle^0_1$      ~& $|(bb)^{\bar{3}}_1\{\bar{b}\bar{c}\}^3_1\rangle^0_1$
                                                                ~& $|(bb)^{\bar{3}}_1\{\bar{b}\bar{c}\}^3_0\rangle^0_1$      \\
                     ~~~~~~~& $2^{+}$                ~~~~~~~& $|(bb)^{\bar{3}}_1\{\bar{b}\bar{c}\}^3_1\rangle^0_2$      ~&
                           $\cdot\cdot\cdot$                     ~&         $\cdot\cdot\cdot$                        \\
\hline
$cc\bar{c}\bar{b}$  ~~~~~~~& $0^{+}$                ~~~~~~~& $|(cc)^6_0\{\bar{c}\bar{b}\}^{\bar{6}}_0\rangle^0_0$      ~& $|(cc)^{\bar{3}}_1\{\bar{c}\bar{b}\}^3_1\rangle^0_0$
                                             ~&            $\cdot\cdot\cdot$                         \\
                     ~~~~~~~& $1^{+}$                ~~~~~~~& $|(cc)^6_0\{\bar{c}\bar{b}\}^{\bar{6}}_1\rangle^0_1$      ~& $|(cc)^{\bar{3}}_1\{\bar{c}\bar{b}\}^3_1\rangle^0_1$
                                                                ~& $|(cc)^{\bar{3}}_1\{\bar{c}\bar{b}\}^3_0\rangle^0_1$      \\
                     ~~~~~~~& $2^{+}$                ~~~~~~~& $|(cc)^{\bar{3}}_1\{\bar{c}\bar{b}\}^3_1\rangle^0_2$      ~&
                    $\cdot\cdot\cdot$                           ~&       $\cdot\cdot\cdot$                                 \\
\hline
$bb\bar{c}\bar{c}$  ~~~~~~~& $0^{+}$                ~~~~~~~& $|\{bb\}^6_0\{\bar{c}\bar{c}\}^{\bar{6}}_0\rangle^0_0$   ~& $|\{bb\}^{\bar{3}}_1\{\bar{c}\bar{c}\}^3_1\rangle^0_0$
                                                ~&           $\cdot\cdot\cdot$                 \\
                     ~~~~~~~& $1^{+}$                ~~~~~~~& $|\{bb\}^{\bar{3}}_1\{\bar{c}\bar{c}\}^3_1\rangle^0_1$   ~&
                          $\cdot\cdot\cdot$                     ~&      $\cdot\cdot\cdot$                             \\
                     ~~~~~~~& $2^{+}$                ~~~~~~~& $|\{bb\}^{\bar{3}}_1\{\bar{c}\bar{c}\}^3_1\rangle^0_2$   ~&
       $\cdot\cdot\cdot$      & $\cdot\cdot\cdot$                                       \\
\hline
$bc\bar{b}\bar{c}$  ~~~~~~~& $0^{++}$             ~~~~~~~& $|(bc)^6_1(\bar{b}\bar{c})^{\bar{6}}_1\rangle^0_0$
                                                                ~& $|(bc)^6_0(\bar{b}\bar{c})^{\bar{6}}_0\rangle^0_0$& $\cdot\cdot\cdot$\\
                                                                ~&
                                                                ~& $|(bc)^{\bar{3}}_1(\bar{b}\bar{c})^3_1\rangle^0_0$
                                                                ~& $|(bc)^{\bar{3}}_0(\bar{b}\bar{c})^3_0\rangle^0_0$ & $\cdot\cdot\cdot$  \\
                     ~~~~~~~& $1^{+-}$             ~~~~~~~& $|(bc)^6_1(\bar{b}\bar{c})^{\bar{6}}_1\rangle^0_1$
                                                                ~& $\frac{1}{\sqrt{2}}|(bc)^6_1(\bar{b}\bar{c})^{\bar{6}}_0\rangle^0_1-|(bc)^6_0(\bar{b}\bar{c})^{\bar{6}}_1\rangle^0_1$& $\cdot\cdot\cdot$\\
                                                                ~&
                                                                ~& $|(bc)^{\bar{3}}_1(\bar{b}\bar{c})^3_1\rangle^0_1$
                                                                ~& $\frac{1}{\sqrt{2}}|(bc)^{\bar{3}}_1(\bar{b}\bar{c})^3_0\rangle^0_1-|(bc)^{\bar{3}}_0(\bar{b}\bar{c})^3_1\rangle^0_1$& $\cdot\cdot\cdot$\\
                     ~~~~~~~& $1^{++}$             ~~~~~~~& $\frac{1}{\sqrt{2}}|(bc)^6_1(\bar{b}\bar{c})^{\bar{6}}_0\rangle^0_1+|(bc)^6_0(\bar{b}\bar{c})^{\bar{6}}_1\rangle^0_1$
                                                                ~& $\frac{1}{\sqrt{2}}|(bc)^{\bar{3}}_1(\bar{b}\bar{c})^3_0\rangle^0_1+|(bc)^{\bar{3}}_0(\bar{b}\bar{c})^3_1\rangle^0_1$
                                                                ~&
                                       $\cdot\cdot\cdot$  \\
                     ~~~~~~~& $2^{++}$             ~~~~~~~& $|(bc)^6_1(\bar{b}\bar{c})^{\bar{6}}_1\rangle^0_2$
                                                                ~& $|(bc)^{\bar{3}}_1(\bar{b}\bar{c})^3_1\rangle^0_2$
                                                                ~&
                                         $\cdot\cdot\cdot$   \\
\hline\hline
\end{tabular}
\end{center}
\end{table*}

\subsubsection{States classified in the quark model}

To calculate the spectroscopy of a $Q_1Q_2\bar{Q}_3\bar{Q}_4$ system,
first we construct the configurations in the product space of spatial~$\otimes$~flavor~$\otimes$~color~$\otimes$~spin.
In the flavor space, the available configurations for all all-heavy tetraquark systems with different flavors
are $bb\bar{b}\bar{c}$, $cc\bar{c}\bar{b}$, $bb\bar{c}\bar{c}$, and $bc\bar{b}\bar{c}$.
This implies that the flavor wave function is symmetric under the exchange of two identical quarks~(antiquarks).
Note that three additional $bc\bar{b}\bar{b}$, $bc\bar{c}\bar{c}$, and $cc\bar{b}\bar{b}$ systems
are not included, as they correspond to the antiparticles of $bb\bar{b}\bar{c}$, $cc\bar{c}\bar{b}$, and $bb\bar{c}\bar{c}$, respectively.

For a tetraquark system, six spin configurations~\big($\chi^{S_{12}S_{34}}_{SS_z}$\big)
and two colorless configurations~\big($|6_{12}\bar{6}_{34}\rangle_c$ and $|\bar{3}_{12}3_{34}\rangle_c$\big)
can be constructed in the spin and color spaces based on SU(2) and SU(3) group representation theories, respectively.
$S_{12}$ stands for the spin quantum number of the diquark ($Q_1Q_2$), while $S_{34}$ stands for that of the other antidiquark ($\bar{Q}_3\bar{Q}_4$).
$S$ is the total spin quantum number of the tetraquark system
while $S_z$ stands for the third component of the total spin $\boldsymbol{S}$.
The explicit forms of the six spin configurations and two colorless configurations can be found in Ref.~\cite{ms100:2019}.

In the spatial space, the relative Jacobi coordinates with
the single-particle coordinates $\boldsymbol{r}_i$~($i=1,2,3,4$) are defined by
\begin{eqnarray}
\begin{pmatrix}
\vxi_1 \\[2ex]
\vxi_2 \\[2ex]
\vxi_3 \\[2ex]
\boldsymbol{R}
\end{pmatrix}
=
\begin{pmatrix}
1 & -1 & 0 & 0 \\[1.5ex]
0 & 0 & 1 & -1 \\[1.5ex]
\dfrac{m_1}{m_{12}} & \dfrac{m_2}{m_{12}} & -\dfrac{m_3}{m_{34}} & -\dfrac{m_4}{m_{34}} \\[2ex]
\dfrac{m_1}{M} & \dfrac{m_2}{M} & \dfrac{m_3}{M} & \dfrac{m_4}{M}
\end{pmatrix}
\begin{pmatrix}
\boldsymbol{r}_1 \\[2ex]
\boldsymbol{r}_2 \\[2ex]
\boldsymbol{r}_3 \\[2ex]
\boldsymbol{r}_4
\end{pmatrix},
\end{eqnarray}
where $m_{ij}=m_i+m_j$ and $M=\sum_{i=1}^4m_i$.
Using the above Jacobi coordinates, it is easy to obtain basis functions
that have well-defined symmetry under permutations of the identical (anti)quark pairs~\cite{Vijande:2009kj}.
For the low-lying $1S$ states under focus in this work, there is no excitation between identical (anti)quarks, the spatial wave functions are constructed to be symmetric under the exchange of the identical (anti)diquark.
It should be noted that for the low-lying $1S$ states, the orbital angular momentum between non-identical (anti)quarks is not necessarily zero (the reason will be discussed in Sec.~\ref{Framework}(A3) below). Therefore, there is no constraint on the symmetry of the spatial wave function under the exchange of two non-identical (anti)quarks.

Finally, considering the Pauli principle, the numbers of $1S$ configurations are:
6 for both the $bb\bar{b}\bar{c}$ and $cc\bar{c}\bar{b}$ systems, 4 for $bb\bar{c}\bar{c}$, and 12 for $bc\bar{b}\bar{c}$.
It should be pointed out that for the purely neutral $bc\bar{b}\bar{c}$ system, each configuration must be an eigenstate under charge conjugation.
All these $1S$-wave configurations for the $bb\bar{b}\bar{c}$, $cc\bar{c}\bar{b}$, $bb\bar{c}\bar{c}$, and $bc\bar{b}\bar{c}$ systems are given in Table~\ref{configurations}.

\begin{table}[h]
\begin{center}
\caption{Explicit forms of the variational parameters $\nu_{ij}$ for each system}
\label{Variational Parameters}
\begin{tabular}{cccccccccccccc} \hline \hline
&~~~                    &~~~$\nu_{12}$ ~~~&$\nu_{34}$ ~~~&$\nu_{13}$ ~~~&$\nu_{24}$ ~~~&$\nu_{14}$ ~~~&$\nu_{23}$~~~\\
\hline
&~~~$bb\bar{b}\bar{c}$  &~~~$a$ ~~~&$d$ ~~~&$p$ ~~~&$e$ ~~~&$e$ ~~~&$p$~~~\\
&~~~$cc\bar{c}\bar{b}$  &~~~$a$ ~~~&$d$ ~~~&$p$ ~~~&$e$ ~~~&$e$ ~~~&$p$~~~\\
&~~~$bb\bar{c}\bar{c}$  &~~~$a$ ~~~&$e$ ~~~&$p$ ~~~&$p$ ~~~&$p$ ~~~&$p$~~~\\
&~~~$bc\bar{b}\bar{c}$  &~~~$a$ ~~~&$a$ ~~~&$e$ ~~~&$d$ ~~~&$b$ ~~~&$b$~~~\\
\hline \hline
\end{tabular}
\end{center}
\end{table}

\subsubsection{Numerical method}\label{Numerical method}

To solve the four-body problem accurately,
we adopt the explicitly correlated Gaussian (ECG) method~\cite{Varga:1995dm,Varga:1997xga,Mitroy:2013eom}.
It is a well-established variational method to solve quantum few-body problems.
The spatial part of the wave function for the $1S$-wave tetraquark system is expanded in terms of ECG basis set.
Such a basis function can be expressed as
\begin{eqnarray}\label{basis}
\label{breit}
\psi(\boldsymbol{r}_1, \boldsymbol{r}_2, \boldsymbol{r}_3, \boldsymbol{r}_4)
=\exp\left[-\sum_{i<j=1}^4\nu_{ij}(\boldsymbol{r}_i-\boldsymbol{r}_j)^2\right],
\end{eqnarray}
where $\nu_{ij}$ are variational parameters.
Due to the symmetry of identical (anti)quarks, the explicit expressions of the variational parameters $\nu_{ij}$
for the different systems are provided in Table~\ref{Variational Parameters}.

It is convenient to use a set of the Jacobi coordinates $\vxi=(\vxi_1, \vxi_2, \vxi_3)$,
instead of the relative distance vectors $\boldsymbol{r}_{ij}$.
Then the correlated Gaussian basis function can be rewritten as
\begin{eqnarray}\label{basis1}
G(\vxi, \mathbb{A})=\mathrm{exp}\left(-\sum_{i,j}A_{ij}~\vxi_i\cdot\vxi_j\right)\equiv \mathrm{exp}(-\vxi^T\mathbb{A}~\vxi),
\end{eqnarray}
where $\mathbb{A}$ is a $3\times3$ matrix, which is related to the variational parameters.
Since the definition of Jacobi coordinates is not unique,
we can also choose two alternative sets of Jacobi coordinates, denoted as $\vxi'$ and $\vxi''$, i.e.,
\begin{eqnarray}
\begin{pmatrix}
\vxi'_1 \\[2ex]
\vxi'_2 \\[2ex]
\vxi'_3 \\[2ex]
\boldsymbol{R}
\end{pmatrix}
=
\begin{pmatrix}
1 & 0 & -1 & 0 \\[1.5ex]
0 & 1 & 0 & -1 \\[1.5ex]
\dfrac{m_1}{m_{13}} & -\dfrac{m_2}{m_{24}} & \dfrac{m_3}{m_{13}} & -\dfrac{m_4}{m_{24}} \\[2ex]
\dfrac{m_1}{M} & \dfrac{m_2}{M} & \dfrac{m_3}{M} & \dfrac{m_4}{M}
\end{pmatrix}
\begin{pmatrix}
\boldsymbol{r}_1 \\[2ex]
\boldsymbol{r}_2 \\[2ex]
\boldsymbol{r}_3 \\[2ex]
\boldsymbol{r}_4
\end{pmatrix},
\end{eqnarray}
and
\begin{eqnarray}
\begin{pmatrix}
\vxi''_1 \\[2ex]
\vxi''_2 \\[2ex]
\vxi''_3 \\[2ex]
\boldsymbol{R}
\end{pmatrix}
=
\begin{pmatrix}
1 & 0 & 0 & -1 \\[1.5ex]
0 & 1 & -1 & 0 \\[1.5ex]
\dfrac{m_1}{m_{14}} & -\dfrac{m_2}{m_{23}} & -\dfrac{m_3}{m_{23}} & \dfrac{m_4}{m_{14}} \\[2ex]
\dfrac{m_1}{M} & \dfrac{m_2}{M} & \dfrac{m_3}{M} & \dfrac{m_4}{M}
\end{pmatrix}
\begin{pmatrix}
\boldsymbol{r}_1 \\[2ex]
\boldsymbol{r}_2 \\[2ex]
\boldsymbol{r}_3 \\[2ex]
\boldsymbol{r}_4
\end{pmatrix}.
\end{eqnarray}
The coordinates $\vxi'$ or $\vxi''$ are convenient in describing the direct and exchange meson-meson channels.
Using the Jacobi coordinates $\vxi'$ and $\vxi''$ instead of the relative distance vectors $\boldsymbol{r}_{ij}$,
the correlated Gaussian basis function can also be rewritten as
\begin{eqnarray}\label{basis2}
G(\vxi', \mathbb{A}')=\mathrm{exp}\left(-\sum_{i,j}A'_{ij}~\vxi'_i\cdot\vxi'_j\right)\equiv \mathrm{exp}(-\vxi'^T\mathbb{A}'~\vxi'),
\end{eqnarray}
and
\begin{eqnarray}\label{basis2}
G(\vxi'', \mathbb{A}'')=\mathrm{exp}\left(-\sum_{i,j}A''_{ij}~\vxi''_i\cdot\vxi''_j\right)\equiv \mathrm{exp}(-\vxi''^T\mathbb{A}''~\vxi'').
\end{eqnarray}
Since the three sets of basis functions $G(\vxi, \mathbb{A})$, $G(\vxi', \mathbb{A}')$, and $G(\vxi'', \mathbb{A}'')$
obtained via different Jacobi coordinate transformations
are all derived from the same parent function
$\psi(\boldsymbol{r}_1, \boldsymbol{r}_2, \boldsymbol{r}_3, \boldsymbol{r}_4)$,
they are completely equivalent~\cite{Brink:1998as}, i.e.,
\begin{eqnarray}\label{basis2}
G(\vxi, \mathbb{A})=G(\vxi', \mathbb{A}')=G(\vxi'', \mathbb{A}'').
\end{eqnarray}
This indicates that if the form of the basis function $G(\vxi, \mathbb{A})$ is ensured to be complete,
it is feasible to calculate the mass spectrum using only one set of Jacobi coordinates $\vxi$.

In the following, we will perform a detailed analysis of the correlated Gaussian basis $G(\vxi, \mathbb{A})$.
To illustrate this basis, we first take the $bc\bar{b}\bar{c}$ system as an example.
The matrix $\mathbb{A}$ in Eq.~\eqref{basis1} can be written explicitly for the $bc\bar{b}\bar{c}$ system as
\begin{widetext}
\begin{equation}\label{matrix 1}
\mathbb{A} =
\begin{pmatrix}
a + \dfrac{(p + e) m_c^2 + (p + d) m_b^2}{(m_b + m_c)^2} &
\dfrac{2p m_b m_c - e m_c^2 - d m_b^2}{(m_b + m_c)^2} &
\dfrac{(p + e) m_c - (p + d) m_b}{m_b + m_c} \\[1.2em]
\dfrac{2p m_b m_c - e m_c^2 - d m_b^2}{(m_b + m_c)^2} &
a + \dfrac{(p + e) m_c^2 + (p + d) m_b^2}{(m_b + m_c)^2} &
\dfrac{-(p + e) m_c + (p + d) m_b}{m_b + m_c} \\[1.2em]
\dfrac{(p + e) m_c - (p + d) m_b}{m_b + m_c} &
\dfrac{-(p + e) m_c + (p + d) m_b}{m_b + m_c} &
2p + e + d
\end{pmatrix}.
\end{equation}
\end{widetext}
From Eq.~\eqref{matrix 1}, one can see that the matrix $\mathbb{A}$ above contains four independent variational parameters $a$, $p$, $e$, $d$, and has non-zero off-diagonal elements. This complex structure contrasts sharply with the simplified form used in our previous work~\cite{ms100:2019}, where the matrix $\mathbb{A}$ for the $bc\bar{b}\bar{c}$ system was written explicitly as
\begin{equation}\label{matrix 2}
\mathbb{A} =
\begin{pmatrix}
\dfrac{m_b m_c}{2(m_b + m_c)} \omega_\ell & 0 & 0 \\[1.2em]
0 & \dfrac{m_b m_c}{2(m_b + m_c)} \omega_\ell & 0 \\[1.2em]
0 & 0 & \dfrac{m_b + m_c}{4} \omega_\ell
\end{pmatrix}.
\end{equation}
This matrix contains only one independent variational parameter $\omega_\ell$ and has no non-zero off-diagonal elements, reflecting the incompleteness of the trial wave function adopted in our previous work~\cite{ms100:2019}.
Furthermore, we focus on the off-diagonal terms in the matrix $\mathbb{A}$.
For example, in Eq.~\eqref{matrix 1}, the off-diagonal term $A_{12}$ ($=A_{21}$) is nonzero,
indicating the presence of a cross term exp$({-2A_{12}~\vxi_1\cdot\vxi_2})$ in the basis functions.
One can perform a partial-wave expansion on the cross term:
\begin{eqnarray}\label{partial-wave expansion}
e^{-2A_{12}\,\vxi_1\cdot\vxi_2}
= 4\pi \sum_{l=0}^\infty \sum_{m=-l}^{l}
i_l(-2A_{12}\xi_1\xi_2) \,
Y^*_{lm}(\hat{\vxi}_1) Y_{lm}(\hat{\vxi}_2),
\end{eqnarray}
where $i_l(z)$ is the modified spherical Bessel function of the first kind.
$Y_{lm}(\hat{\vxi}_i)$ is the spherical harmonic function,
where $l$ and $m$ are the quantum numbers of the orbital
angular momentum and its $z$-component corresponding
to the $\vxi_i$-mode excitation, respectively.
According to Eq.~\eqref{partial-wave expansion}, for the low-lying $1S$-wave state of the $bc\bar{b}\bar{c}$ system,
$l_{\xi_1} = l_{\xi_2} = 0$ is only one of its components, it also has $l_{\xi_1} = l_{\xi_2} \neq 0$ components.
The additional contributions from higher partial waves arise from the cross term,
which exists because the $b$ and $c$ quarks and the two antiquarks in the $bc\bar{b}\bar{c}$ system are nonidentical.
Subsequently, we take the $bb\bar{b}\bar{c}$ system as an example to discuss the case where identical quarks or identical antiquarks are present.
The matrix $\mathbb{A}$ can be written explicitly for the $bb\bar{b}\bar{c}$ system as
\begin{eqnarray}
\mathbb{A} =
\begin{pmatrix}
a + \dfrac{p+e}{2} & 0 & 0 \\
0 & d + \dfrac{2(p m_c^2 + e m_b^2)}{(m_b+m_c)^2} & \dfrac{2(e m_b - p m_c)}{m_b+m_c} \\
0 & \dfrac{2(e m_b - p m_c)}{m_b+m_c} & 2(p+e)
\end{pmatrix}.
\end{eqnarray}
In contrast to the $bc\bar{b}\bar{c}$ case, in the matrix above,
$A_{12} = A_{21} = 0$ and $A_{13} = A_{31} = 0$.
This indicates that $\vxi_1$ does not appear in the cross terms of the basis functions,
which is due to the fact that in the $bb\bar{b}\bar{c}$ system the two $b$ quarks are identical.
Therefore, for the low-lying $1S$ $bb\bar{b}\bar{c}$ system,
the relative angular momentum between the two identical quarks
has no contribution from higher partial waves, i.e., only $l_{\xi_1} = 0$.
This indicates that under the exchange of the two identical quarks, the spatial wave function has a definite symmetry.
In summary, by all accounting for the non-identical nature between (anti)quarks,
the trial Gaussian basis functions used in this work contain more independent variational parameters and cross terms compared with the previous work~\cite{ms100:2019},
which lets the basis functions become more complete.

The spatial part of the trial wave function $\Psi(\xi,\mathbb{A})$ can be formed as a linear combination of the correlated Gaussians
\begin{eqnarray}
\Psi(\vxi, \mathbb{A})=\sum_{k=1}^{N}c_kG(\vxi, \mathbb{A}_k).
\end{eqnarray}
The accuracy of the trial function depends on the length of the
expansion $N$ and the nonlinear parameters $c_k$.
In our calculations, following the method of Ref.~\cite{Hiyama:2003cu},
we let the variational parameters form a geometric progression.
For example, for a variational parameter $a$, we take
\begin{eqnarray}
a_i=\frac{1}{2(a_1q^{i-1})^2}~~~~~~(i=1, \cdot\cdot\cdot, n^a_{max}).
\end{eqnarray}
The Gaussian size parameters $\{a_1, a_{n^a_{max}}, n^a_{max}\}$
will be determined through the variation method.
In the calculations, the final results should be stable
and independent with these parameters.

For a given tetraquark configuration, one can work out the Hamiltonian matrix elements,
\begin{eqnarray}
H_{kk'}=\langle\psi_{CS}G(\vxi, \mathbb{A}_k)|H|\psi_{CS}G(\vxi, \mathbb{A}_k')\rangle,
\end{eqnarray}
where $\psi_{CS}$ is the spin-color wave function.
Then, by solving the generalized matrix eigenvalue problem,
\begin{eqnarray}
\sum^N_{k'=1}(H_{kk'}-EN_{kk'})c_{k'}=0,
\end{eqnarray}
one can obtain the eigenenergy $E$, and the expansion coefficients $\{c_k\}$.
The $N_{kk'}$ is an overlap factor defined by $N_{kk'}=\langle G(\vxi, \mathbb{A}_k)|G(\vxi, \mathbb{A}_{k'})\rangle$.

\begin{figure*}[htbp]
 \centering \epsfxsize=18.2 cm \epsfbox{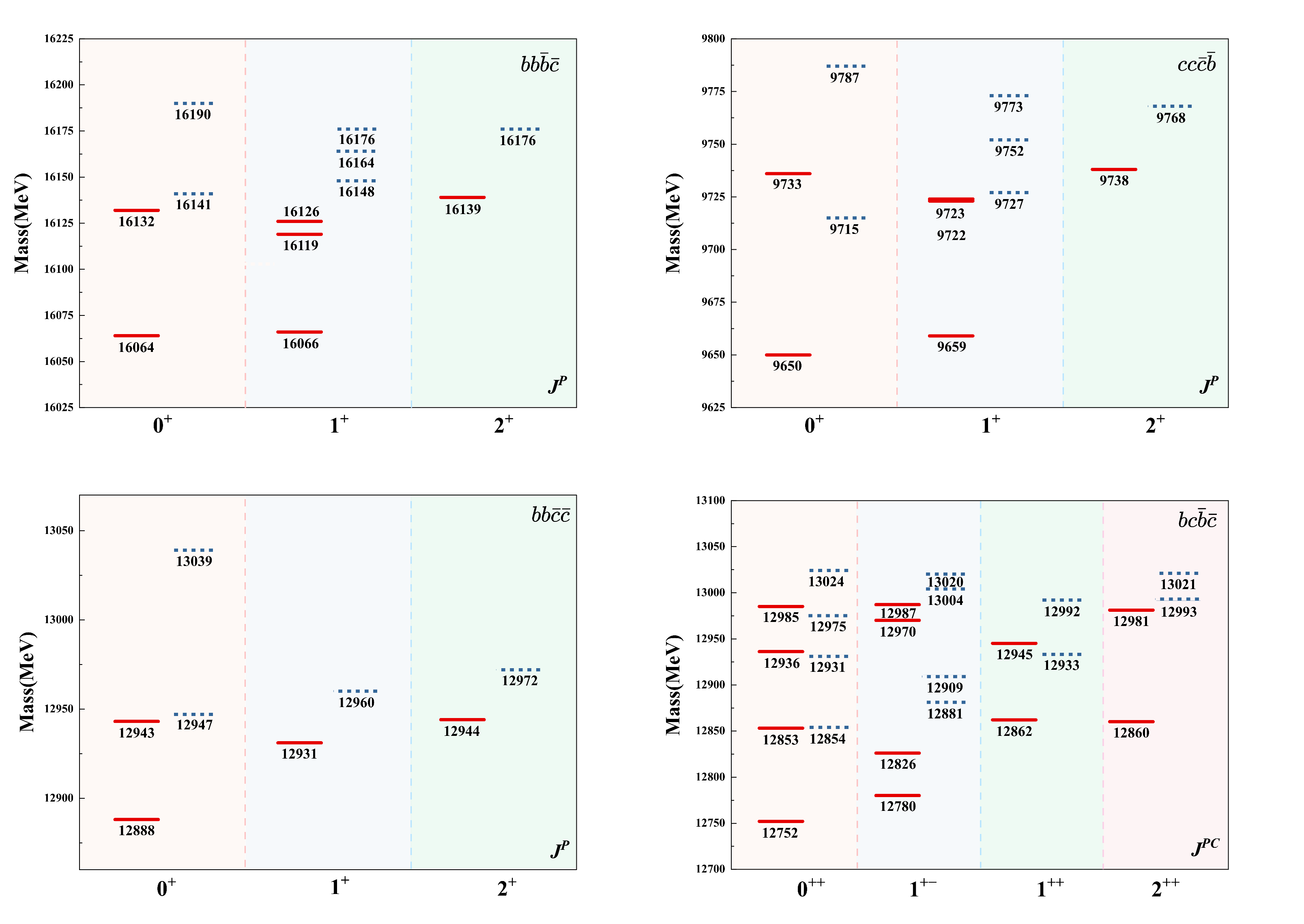}
 \caption{Mass spectrum of all-heavy tetraquarks with different flavors.
 The red solid lines and blue dashed lines represent the results of this work and our previous work~\cite{ms100:2019}, respectively.
 The unit of mass is MeV.}\label{mass spectrum figure}
\end{figure*}

\subsection{Fall-apart decay}

In this work, we calculate the fall-apart decays of the all-heavy tetraquarks with different flavors in a quark-exchange model~\cite{Barnes:1991em,Barnes:2000hu}.
Recently, this model has also been successfully extended to study the fall-apart decays of tetraquarks~\cite{liu:2020eha,Liu:2022hbk,Xiao:2019spy,Wang:2020prk,Han:2022fup,Liu:2024fnh,Liu:2026ljb},
pentaquarks~\cite{Dong:2020nwk,Wang:2019spc,Liang:2024met,An:2025qfw}, and hexaquark states~\cite{An:2025rjv}.
In this model, the quark-quark and quark-antiquark interactions $V_{ij}$ are considered to be
the sources of the fall-apart decays of multiquark states via the quark rearrangement.

For the decay process $A \to BC$, the decay amplitude $\mathcal{M}(A \to BC)$ is described by
\begin{eqnarray}
\label{breit}
\mathcal{M}(A \to BC) = -\sqrt{(2\pi)^3} \sqrt{8 M_A E_B E_C} \left\langle BC \big| \sum_{i<j} V_{ij} \big| A \right\rangle,
\end{eqnarray}
where $A$ stands for the initial tetraquark state, and $BC$ stands for the final hadron pair.
$M_A$ is the mass of the initial state,
while $E_B$ and $E_C$ are the energies of the final states $B$ and $C$, respectively,
in the initial-hadron-rest system.
While $V_{ij}$ stands for the interactions between the inner quarks of final
hadrons $B$ and $C$ (note that $ij= 13,24$ or $ij= 14,23$),
they are taken the same as that of the potential model given in Eq.~(\ref{OGE}).
Then, the partial decay width of the $A \to BC$ process is given by
\begin{eqnarray}
\label{breit2}
\Gamma = \frac{1}{s!}\frac{1}{2 J_A + 1} \frac{|\boldsymbol{q}|}{8 \pi M_A^2} \left| \mathcal{M}(A \to BC) \right|^2,
\end{eqnarray}
where $\boldsymbol{q}$ is the three-vector momentum of the final state $B$
or $C$ in the initial-hadron-rest frame.
The term $\frac{1}{s!}$ represents a statistical factor that accounts for the indistinguishability of particles.
In scenarios where the final state contains two or more identical particles,
it is necessary to divide by the number of permutations among these particles to avoid overcounting, as they are indistinguishable from one another.

\begin{table}[h]
\begin{center}
\caption{Masses, root-mean-square radii, and effective
harmonic oscillator parameters $\beta$ for the final meson states involving
in the rearrangement decays.}
\label{meson mass}
\begin{tabular}{ccccccccccc} \hline \hline
~~~&State           ~~~~~~&$J^P$   ~~~~~~&Mass~(MeV)                               ~~~~~~&$\sqrt{\langle r^2 \rangle}$~(fm)     ~~~~~~&$\beta$~(GeV)~~~\\
\hline
~~~&$\eta_c$        ~~~~~~&$0^-$   ~~~~~~&2984~\cite{ParticleDataGroup:2024cfk}    ~~~~~~&0.363                                 ~~~~~~&0.665~~~\\
~~~&$J/\psi$        ~~~~~~&$1^-$   ~~~~~~&3097~\cite{ParticleDataGroup:2024cfk}    ~~~~~~&0.415                                 ~~~~~~&0.583~~~\\
\\
~~~&$\eta_b$        ~~~~~~&$0^-$   ~~~~~~&9399~\cite{ParticleDataGroup:2024cfk}    ~~~~~~&0.196                                 ~~~~~~&1.231~~~\\
~~~&$\Upsilon(1S)$  ~~~~~~&$1^-$   ~~~~~~&9460~\cite{ParticleDataGroup:2024cfk}    ~~~~~~&0.212                                 ~~~~~~&1.139~~~\\
\\
~~~&$B_c$           ~~~~~~&$0^-$   ~~~~~~&6274~\cite{ParticleDataGroup:2024cfk}    ~~~~~~&0.306                                 ~~~~~~&0.791~~~\\
~~~&$B_c^*$         ~~~~~~&$1^-$   ~~~~~~&6328                                     ~~~~~~&0.327                                 ~~~~~~&0.740~~~\\
\hline \hline
\end{tabular}
\end{center}
\end{table}

In the present work, the masses and wave functions of the initial tetraquark states
are the numerical results obtained from our potential model calculations.
For the final mesons $B$ and $C$, their wave functions are approximated by a single harmonic oscillator (SHO) form, i.e., $e^{-\frac{1}{2}\beta^2 r^2}$ for simplicity.
Their SHO parameters $\beta$ are determined by fitting the root mean square radii,
which are obtained from our potential model calculations with the same Hamiltonian given in Eq.~(\ref{Hamiltonian}).
Our determined root-mean-square (RMS) radii and SHO parameters for the final meson states are collected in Table~\ref{meson mass}.
For the unestablished $B_c^{*}$ in the final state, the mass is adopted from our quark model predictions with Eq.~(\ref{Hamiltonian}),
while for the well-established meson states, the masses are taken from the PDG averaged values~\cite{ParticleDataGroup:2024cfk}.
The masses for the final meson states are collected in Table~\ref{meson mass} as well.

\begin{table*}[htbp]
\centering
\renewcommand{\arraystretch}{1.4}
\caption{\label{numerical results of the mass spectrum}The numerical results of the mass spectrum~(in MeV), the mass contributions of each Hamiltonian part~(in MeV), and the root-mean-square radii~(in fm) for the 1$S$-wave eigenstates of the $bb\bar{b}\bar{c}$, $cc\bar{c}\bar{b}$, $bb\bar{c}\bar{c}$, and $bc\bar{b}\bar{c}$ systems. In the table, we define that $R_{ij}=\sqrt{\langle
r_{ij}^2 \rangle}$,
$\Big|(bc)^6_1(\bar{b}\bar{c})^{\bar{6}}_0\Big\rangle_1^{0\pm} \equiv \frac{1}{\sqrt{2}}\Big( |(bc)^6_1(\bar{b}\bar{c})^{\bar{6}}_0\rangle_1^0 \pm |(bc)^6_0(\bar{b}\bar{c})^{\bar{6}}_1\rangle_1^0 \Big)$,
and $\Big|(bc)^{\bar{3}}_1(\bar{b}\bar{c})^3_0\Big\rangle_1^{0\pm} \equiv \frac{1}{\sqrt{2}}\Big( |(bc)^{\bar{3}}_1(\bar{b}\bar{c})^3_0\rangle_1^0 \pm |(bc)^{\bar{3}}_0(\bar{b}\bar{c})^3_1\rangle_1^0 \Big)$.}
\begin{ruledtabular}
\begin{tabular}{ccccccc}
    & $J^{P(C)}$ & Eigenstate & Mass & $\langle T\rangle/\langle V^{Conf}\rangle/\langle V^{Coul}\rangle/\langle V^{SS}\rangle$ & $R_{12}/R_{34}/R_{13}/R_{24}/R_{14}/R_{23}$ \\
    \hline
    \multirow{8}{*}{$bb\bar{b}\bar{c}$}
    & \multirow{2}{*}{$0^{+}$}
    & \multirow{2}{*}{$\begin{pmatrix} -0.43 & 0.90 \\ -0.90 & -0.43 \end{pmatrix}$ $\begin{pmatrix} \ket{\{bb\}^6_0(\bar{b}\bar{c})^{\bar{6}}_0}_0^0 \\ \ket{\{bb\}^{\bar{3}}_1(\bar{b}\bar{c})^3_1}_0^0 \end{pmatrix}$}
    & \multirow{2}{*}{$\begin{pmatrix} 16132 \\ 16064 \end{pmatrix}$}
    & $779/424/{-1129}/{19}$ & $0.28/0.38/0.29/0.38/0.38/0.29$ \\
    &&&& $777/416/{-1153}/{-15}$ & $0.32/0.40/0.28/0.38/0.38/0.28$ \\
    \\
    & \multirow{3}{*}{$1^{+}$}
    & \multirow{3}{*}{$\begin{pmatrix} 0.18 & 0.14 & 0.99 \\ 0.29 & -0.95 & 0.13 \\ -0.96 & -0.29 & 0.06 \end{pmatrix}$ $\begin{pmatrix} \ket{\{bb\}^6_0(\bar{b}\bar{c})^{\bar{6}}_1}_1^0 \\ \ket{\{bb\}^{\bar{3}}_1(\bar{b}\bar{c})^3_0}_1^0 \\ \ket{\{bb\}^{\bar{3}}_1(\bar{b}\bar{c})^3_1}_1^0 \end{pmatrix}$}
    & \multirow{3}{*}{$\begin{pmatrix} 16126 \\ 16119 \\ 16066 \end{pmatrix}$}
    & $800/444/{-1159}/2$ & $0.27/0.37/0.30/0.38/0.38/0.30$ \\
    &&&& $792/427/{-1138}/{-1}$ & $0.28/0.37/0.30/0.38/0.38/0.30$ \\
    &&&& $797/418/{-1180}/{-8}$ & $0.32/0.40/0.28/0.38/0.38/0.28$ \\
    \\
    & $2^{+}$
    & $\ket{\{bb\}^{\bar{3}}_1(\bar{b}\bar{c})^3_1}_2^0$
    & $16139$
    & $755/437/{-1106}/14$
    & $0.27/0.37/0.30/0.39/0.39/0.30$ \\

    \hline\hline\\

    \multirow{8}{*}{$cc\bar{c}\bar{b}$}
    & \multirow{2}{*}{$0^{+}$}
    & \multirow{2}{*}{$\begin{pmatrix} -0.55 & 0.84 \\ -0.84 & -0.55 \end{pmatrix}$ $\begin{pmatrix} \ket{\{cc\}^6_0(\bar{c}\bar{b})^{\bar{6}}_0}_0^0 \\ \ket{\{cc\}^{\bar{3}}_1\{\bar{c}\bar{b}\}^3_1}_0^0 \end{pmatrix}$}
    & \multirow{2}{*}{$\begin{pmatrix} 9733 \\ 9650 \end{pmatrix}$}
    & $772/534/{-935}/64$ & $0.48/0.42/0.48/0.42/0.42/0.48$ \\
    &&&& $780/468/{-942}/43$ & $0.50/0.46/0.48/0.41/0.41/0.48$ \\
    \\
    & \multirow{3}{*}{$1^{+}$}
    & \multirow{3}{*}{$\begin{pmatrix} 0.37 & 0.92 & 0.12 \\ 0.25 & -0.03 & 0.97 \\ 0.89 & -0.39 & 0.22 \end{pmatrix}$ $\begin{pmatrix} \ket{\{cc\}^6_0(\bar{c}\bar{b})^{\bar{6}}_1}_1^0 \\ \ket{\{cc\}^{\bar{3}}_1(\bar{c}\bar{b})^3_0}_1^0 \\ \ket{\{cc\}^{\bar{3}}_1(\bar{c}\bar{b})^3_1}_1^0 \end{pmatrix}$}
    & \multirow{3}{*}{$\begin{pmatrix} 9723 \\ 9722 \\ 9659 \end{pmatrix}$}
    & $751/578/{-920}/13$ & $0.47/0.39/0.49/0.43/0.43/0.49$ \\
    &&&& $750/584/{-920}/7$ & $0.48/0.41/0.49/0.43/0.43/0.49$ \\
    &&&& $740/562/{-931}/{-13}$ & $0.51/0.46/0.48/0.41/0.41/0.48$ \\
    \\
    & $2^{+}$
    & $\ket{\{cc\}^{\bar{3}}_1(\bar{c}\bar{b})^3_1}_2^0$
    & $9738$
    & $721/593/{-898}/20$
    & $0.47/0.40/0.50/0.44/0.44/0.50$ \\

    \hline\hline\\

    \multirow{6}{*}{$bb\bar{c}\bar{c}$}
    & \multirow{2}{*}{$0^{+}$}
    & \multirow{2}{*}{$\begin{pmatrix} -0.57 & 0.82 \\ -0.82 & -0.57 \end{pmatrix}$ $
    \begin{pmatrix} \ket{\{bb\}^6_0\{\bar{c}\bar{c}\}^{\bar{6}}_0}_0^0
    \\\ket{\{bb\}^{\bar{3}}_1\{\bar{c}\bar{c}\}^3_1}_0^0 \end{pmatrix}$}
    & \multirow{2}{*}{$\begin{pmatrix} 12942 \\ 12888 \end{pmatrix}$}
    & $752/503/{-1011}/29$ & $0.32/0.47/0.40/0.40/0.40/0.40$ \\
    &&&& $747/500/{-1013}/{-16}$ & $0.35/0.48/0.40/0.40/0.40/0.40$ \\
    \\
    & $1^{+}$
    & $\ket{\{bb\}^{\bar{3}}_1\{\bar{c}\bar{c}\}^3_1}_1^0$
    & $12931$
    & $760/507/{-1012}/{-1}$
    & $0.28/0.46/0.41/0.41/0.41/0.41$ \\
    \\
    & $2^{+}$
    & $\ket{\{bb\}^{\bar{3}}_1\{\bar{c}\bar{c}\}^3_1}_2^0$
    & $12944$
    & $734/516/{-994}/18$
    & $0.29/0.46/0.41/0.41/0.41/0.41$ \\

    \hline\hline\\

    \multirow{12}{*}{$bc\bar{b}\bar{c}$}
    & \multirow{4}{*}{$0^{++}$}
    & \multirow{4}{*}{$\begin{pmatrix} -0.38 & -0.34 & 0.47 & 0.72 \\ -0.29 & 0.12 & 0.75 & -0.59 \\ 0.81 & -0.46 & 0.36 & -0.03 \\ -0.35 & -0.81 & -0.30 & -0.38 \end{pmatrix}$ $\begin{pmatrix} \ket{(bc)^6_0(\bar{b}\bar{c})^{\bar{6}}_0}_0^0 \\ \ket{(bc)^6_1(\bar{b}\bar{c})^{\bar{6}}_1}_0^0 \\ \ket{(bc)^{\bar{3}}_0(\bar{b}\bar{c})^3_0}_0^0 \\ \ket{(bc)^{\bar{3}}_1(\bar{b}\bar{c})^3_1}_0^0 \end{pmatrix}$}
    & \multirow{4}{*}{$\begin{pmatrix} 12985 \\ 12936 \\ 12853 \\ 12752 \end{pmatrix}$}
    & $779/510/{-989}/15$ & $0.37/0.37/0.31/0.46/0.40/0.40$ \\
    &&&& $775/509/{-986}/{-31}$ & $0.34/0.34/0.31/0.46/0.40/0.40$ \\
    &&&& $791/477/{-1100}/{14}$ & $0.41/0.41/0.30/0.46/0.40/0.40$ \\
    &&&& $829/465/{-1130}/{-84}$ & $0.41/0.41/0.30/0.46/0.41/0.41$ \\
    \\
    & \multirow{4}{*}{$1^{+-}$}
    & \multirow{4}{*}{$\begin{pmatrix} -0.08 & 0.14 & -0.53 & 0.84 \\ 0.39 & -0.91 & -0.06 & 0.15 \\ -0.67 & -0.27 & 0.59 & 0.35 \\ -0.62 & -0.30 & -0.61 & -0.40 \end{pmatrix}$ $\begin{pmatrix} \ket{(bc)^6_1(\bar{b}\bar{c})^{\bar{6}}_1}_1^0 \\ \ket{(bc)^{\bar{3}}_1(\bar{b}\bar{c})^3_1}_1^0 \\ \ket{(bc)^6_1(\bar{b}\bar{c})^{\bar{6}}_0}_1^{0-} \\ \ket{(bc)^{\bar{3}}_1(\bar{b}\bar{c})^3_0}_1^{0-} \end{pmatrix}$}
    & \multirow{4}{*}{$\begin{pmatrix} 12987 \\ 12970 \\ 12826 \\ 12780 \end{pmatrix}$}
    & $770/515/{-978}/10$ & $0.40/0.40/0.32/0.47/0.41/0.41$ \\
    &&&& $761/515/{-974}/{-18}$ & $0.43/0.43/0.29/0.46/0.40/0.40$ \\
    &&&& $798/473/{-1110}/{-5}$ & $0.40/0.40/0.30/0.45/0.40/0.40$ \\
    &&&& $794/474/{-1108}/{-49}$ & $0.41/0.41/0.31/0.46/0.40/0.40$ \\
    \\
    & \multirow{2}{*}{$1^{++}$}
    & \multirow{2}{*}{$\begin{pmatrix} -0.29 & 0.96 \\ -0.96 & -0.29 \end{pmatrix}$ $\begin{pmatrix} \ket{(bc)^6_1(\bar{b}\bar{c})^{\bar{6}}_0}_1^{0+} \\ \ket{(bc)^{\bar{3}}_1(\bar{b}\bar{c})^3_0}_1^{0+} \end{pmatrix}$}
    & \multirow{2}{*}{$\begin{pmatrix} 12945 \\ 12862 \end{pmatrix}$}
    & $757/513/{-979}/{-16}$ & $0.39/0.39/0.32/0.47/0.40/0.40$ \\
    &&&& $767/485/{-1079}/19$ & $0.43/0.43/0.30/0.47/0.40/0.40$ \\
    \\
    & \multirow{2}{*}{$2^{++}$}
    & \multirow{2}{*}{$\begin{pmatrix} -0.91 & 0.42 \\ -0.42 & -0.91 \end{pmatrix}$ $\begin{pmatrix} \ket{(bc)^6_1(\bar{b}\bar{c})^{\bar{6}}_1}_2^0 \\ \ket{(bc)^{\bar{3}}_1(\bar{b}\bar{c})^3_1}_2 \end{pmatrix}$}
    & \multirow{2}{*}{$\begin{pmatrix} 12981 \\ 12860 \end{pmatrix}$}
    & $730/524/{-957}/13$ & $0.43/0.43/0.30/0.47/0.42/0.42$ \\
    &&&& $747/489/{-1070}/24$ & $0.40/0.40/0.32/0.47/0.47/0.47$ \\
\end{tabular}
\end{ruledtabular}
\end{table*}

\begin{table*}[htbp]
\centering
\renewcommand{\arraystretch}{1.4}
\caption{\label{decay widths}
The predicted partial decay widths of the fall-apart decay processes of the 1$S$ states for
the $bb\bar{b}\bar{c}$, $cc\bar{c}\bar{b}$, $bb\bar{c}\bar{c}$, and $bc\bar{b}\bar{c}$ systems.
Forbidden decay channels are denoted by ``$\cdots$''.
The unit is MeV.}
\begin{tabular*}{\textwidth}{@{\extracolsep{\fill}} cccccccccccc }
\hline\hline
& $J^P$ & Mass & \multicolumn{2}{c}{$\Gamma[\eta_b B_c^-]$} & \multicolumn{2}{c}{$\Gamma[\eta_b B_c^{*-}]$} & \multicolumn{2}{c}{$\Gamma[\Upsilon B_c^-]$} & \multicolumn{2}{c}{$\Gamma[\Upsilon B_c^{*-}]$} & $\Gamma[\text{Sum}]$ \\
\hline
\multirow{8}{*}{$bb\bar{b}\bar{c}$}
& $0^+$ & 16064 & \multicolumn{2}{c}{0.12} & \multicolumn{2}{c}{$\cdots$} & \multicolumn{2}{c}{$\cdots$} & \multicolumn{2}{c}{1.13} & 1.25 \\
&& 16132 & \multicolumn{2}{c}{$<0.01$} & \multicolumn{2}{c}{$\cdots$} & \multicolumn{2}{c}{$\cdots$} & \multicolumn{2}{c}{0.44} & 0.44 \\
&\\
& $1^+$ & 16066 & \multicolumn{2}{c}{$\cdots$} & \multicolumn{2}{c}{0.70} & \multicolumn{2}{c}{0.37} & \multicolumn{2}{c}{0.78} & 1.85 \\
&& 16119 & \multicolumn{2}{c}{$\cdots$} & \multicolumn{2}{c}{0.96} & \multicolumn{2}{c}{0.66} & \multicolumn{2}{c}{1.34} & 2.96 \\
&& 16126 & \multicolumn{2}{c}{$\cdots$} & \multicolumn{2}{c}{0.01} & \multicolumn{2}{c}{0.03} & \multicolumn{2}{c}{0.10} & 0.14 \\
&\\
& $2^+$ & 16139 & \multicolumn{2}{c}{$\cdots$} & \multicolumn{2}{c}{$\cdots$} & \multicolumn{2}{c}{$\cdots$} & \multicolumn{2}{c}{0.86} & 0.86 \\
\hline\hline
& $J^P$ & Mass  & \multicolumn{2}{c}{$\Gamma[\eta_c B_c^+]$} & \multicolumn{2}{c}{$\Gamma[\eta_c B_c^{*+}]$} & \multicolumn{2}{c}{$\Gamma[J/\psi B_c^+]$} & \multicolumn{2}{c}{$\Gamma[J/\psi B_c^{*+}]$} & $\Gamma[\text{Sum}]$ \\
\hline
\multirow{8}{*}{$cc\bar{c}\bar{b}$}
& $0^+$ & 9650 & \multicolumn{2}{c}{1.50} & \multicolumn{2}{c}{$\cdots$} & \multicolumn{2}{c}{$\cdots$} & \multicolumn{2}{c}{0.85} & 2.35 \\
&& 9733 & \multicolumn{2}{c}{0.03} & \multicolumn{2}{c}{$\cdots$} & \multicolumn{2}{c}{$\cdots$} & \multicolumn{2}{c}{2.65} & 2.68 \\
&\\
& $1^+$ & 9659 & \multicolumn{2}{c}{$\cdots$} & \multicolumn{2}{c}{0.14} & \multicolumn{2}{c}{0.20} & \multicolumn{2}{c}{0.12} & 0.46 \\
&& 9722 & \multicolumn{2}{c}{$\cdots$} & \multicolumn{2}{c}{1.08} & \multicolumn{2}{c}{0.01} & \multicolumn{2}{c}{0.05} & 1.14 \\
&& 9723 & \multicolumn{2}{c}{$\cdots$} & \multicolumn{2}{c}{$<0.01$} & \multicolumn{2}{c}{0.60} & \multicolumn{2}{c}{0.23} & 0.83 \\
&\\
& $2^+$ & 9738 & \multicolumn{2}{c}{$\cdots$} & \multicolumn{2}{c}{$\cdots$} & \multicolumn{2}{c}{$\cdots$} & \multicolumn{2}{c}{0.18} & 0.18 \\
\hline\hline
& $J^P$ & Mass  & \multicolumn{2}{c}{$\Gamma[B_c^- B_c^-]$} & \multicolumn{2}{c}{$\Gamma[B_c^- B_c^{*-}]$} & \multicolumn{2}{c}{$\Gamma[B_c^{*-} B_c^{*-}]$} & \multicolumn{2}{c}{$\Gamma[\text{Sum}]$} \\
\hline
\multirow{6}{*}{$bb\bar{c}\bar{c}$}
& $0^+$ & 12888 & \multicolumn{2}{c}{0.37} & \multicolumn{2}{c}{$\cdots$} & \multicolumn{2}{c}{0.14} & \multicolumn{2}{c}{0.51} \\
&& 12942 & \multicolumn{2}{c}{1.12} & \multicolumn{2}{c}{$\cdots$} & \multicolumn{2}{c}{0.87} & \multicolumn{2}{c}{1.99} \\
&\\
& $1^+$ & 12931 & \multicolumn{2}{c}{$\cdots$} & \multicolumn{2}{c}{0.10} & \multicolumn{2}{c}{$\cdots$} & \multicolumn{2}{c}{0.10} \\
&\\
& $2^+$ & 12944 & \multicolumn{2}{c}{$\cdots$} & \multicolumn{2}{c}{$\cdots$} & \multicolumn{2}{c}{1.52} & \multicolumn{2}{c}{1.52} \\
\hline\hline
& $J^{PC}$ & Mass  & $\Gamma[\eta_b\eta_c]$ & $\Gamma[\eta_b J/\psi]$ & $\Gamma[\Upsilon\eta_c]$ & $\Gamma[\Upsilon J/\psi]$ & $\Gamma[B_c^+ B_c^-]$ & \multicolumn{2}{c}{$\Gamma[B_c^+ B_c^{*-}+B_c^- B_c^{*+}]$} & $\Gamma[B_c^{*+} B_c^{*-}]$ & $\Gamma[\text{Sum}]$ \\
\hline
\multirow{15}{*}{$bc\bar{b}\bar{c}$}
& $0^{++}$ & 12752 & 0.37 & $\cdots$ & $\cdots$ & 0.06 & 0.51 & \multicolumn{2}{c}{$\cdots$} & 2.41 & 3.35 \\
&& 12853 & 0.10 & $\cdots$ & $\cdots$ & 0.50 & 0.01 & \multicolumn{2}{c}{$\cdots$} & 0.39 & 1.00 \\
&& 12936 & 1.58 & $\cdots$ & $\cdots$ & $<0.01$ & 0.06 & \multicolumn{2}{c}{$\cdots$} & 1.69 & 3.33 \\
&& 12985 & 0.02 & $\cdots$ & $\cdots$ & 2.01 & 0.60 & \multicolumn{2}{c}{$\cdots$} & 0.07 & 2.70 \\
&\\
& $1^{+-}$ & 12780 & $\cdots$ & $<0.01$ & 0.03 & $\cdots$ & $\cdots$ & \multicolumn{2}{c}{0.21} & 0.14 & 0.38 \\
&& 12826 & $\cdots$ & 0.03 & $<0.01$ & $\cdots$ & $\cdots$ & \multicolumn{2}{c}{0.04} & $<0.01$ & 0.07 \\
&& 12970 & $\cdots$ & 0.02 & 0.05 & $\cdots$ & $\cdots$ & \multicolumn{2}{c}{0.01} & 0.03 & 0.11 \\
&& 12987 & $\cdots$ & 0.06 & 0.03 & $\cdots$ & $\cdots$ & \multicolumn{2}{c}{0.01} & 0.01 & 0.10 \\
&\\
& $1^{++}$ & 12862 & $\cdots$ & $\cdots$ & $\cdots$ & $<0.01$ & $\cdots$ & \multicolumn{2}{c}{$<0.01$} & $\cdots$ & $<0.01$ \\
&& 12945 & $\cdots$ & $\cdots$ & $\cdots$ & $<0.01$ & $\cdots$ & \multicolumn{2}{c}{$<0.01$} & $\cdots$ & $<0.01$ \\
&\\
& $2^{++}$ & 12860 & $\cdots$ & $\cdots$ & $\cdots$ & 0.16 & $\cdots$ & \multicolumn{2}{c}{$\cdots$} & 0.27 & 0.43 \\
&& 12981 & $\cdots$ & $\cdots$ & $\cdots$ & $<0.01$ & $\cdots$ & \multicolumn{2}{c}{$\cdots$} & 0.08 & 0.08 \\
\hline\hline
\end{tabular*}
\end{table*}

\section{Results and discussions}\label{results}

The mass spectra, the mass contributions from each part of the Hamiltonian, and the root-mean-square radii for the 1$S$-wave states of the $bb\bar{b}\bar{c}$, $cc\bar{c}\bar{b}$, $bb\bar{c}\bar{c}$, and $bc\bar{b}\bar{c}$ systems
are presented in Table~\ref{numerical results of the mass spectrum}.
In addition, the mass spectra predicted in this work, as well as those from our previous work~\cite{ms100:2019}, are plotted in Fig.~\ref{mass spectrum figure}. Compared to our previous predictions~\cite{ms100:2019}, it is found that the masses of all states are
significantly shifted downward by about $30-100$ MeV, and the mass splittings are also notably modified.
However, a more remarkable difference is that the main components have changed for some states.
For example, for the two $J^P=0^+$ states of the $bb\bar{b}\bar{c}$ system,
our previous work~\cite{ms100:2019} predicted that the main component of the higher-mass state is $|{bb}_0^6(\bar{b}\bar{c})_0^{\bar{6}}\rangle_0^0$
and that of the lower-mass state is $|{bb}_1^{\bar{3}}(\bar{b}\bar{c})_1^3\rangle_0^0$.
In contrast, in this work, the main component of the higher-mass state is $|{bb}_1^{\bar{3}}(\bar{b}\bar{c})_1^3\rangle_0^0$
and that of the lower-mass state is $|{bb}_0^6(\bar{b}\bar{c})_0^{\bar{6}}\rangle_0^0$.
The main reason for this difference is that the trial wave function adopted in this work
is more complete than that used previously, as discussed in Sec.~\ref{Framework}~(A3).


The fall-apart decay properties of the $1S$-wave states for the tetrquarks $bb\bar{b}\bar{c}$, $cc\bar{c}\bar{b}$, $bb\bar{c}\bar{c}$, and $bc\bar{b}\bar{c}$ are given in Table~\ref{decay widths}.
It is found that the $1S$-wave tetrquarks are likely to be narrow states,
their fall-apart widths are predicted to range from a few tenths to several MeV.
The main reason of the small fall-apart widths obtained in the present work is
that the fall-apart decays are mainly driven by the chromo-magnetic (spin-spin)
interaction $V^{\mathrm{SS}}$~\cite{Liu:2022hbk,Liu:2024fnh}, which is strongly suppressed by the heavy quark masses.
Our predictions of the narrow width nature for the all-heavy tetraquark states are consistent
with the expectations of the real and complex scaling methods~\cite{21Wu:2024hrv,11Hu:2022zdh}.
However, the coupled-channel calculation of Ref.~\cite{27Ortega:2025lmo} gives broad widths,
because the tetraquark states are treated there as meson-meson molecular states, which leads to strong overlap with the continuum.


\subsection{$bb\bar{b}\bar{c}$}

\begin{figure}[htbp]
 \centering \epsfxsize=9 cm \epsfbox{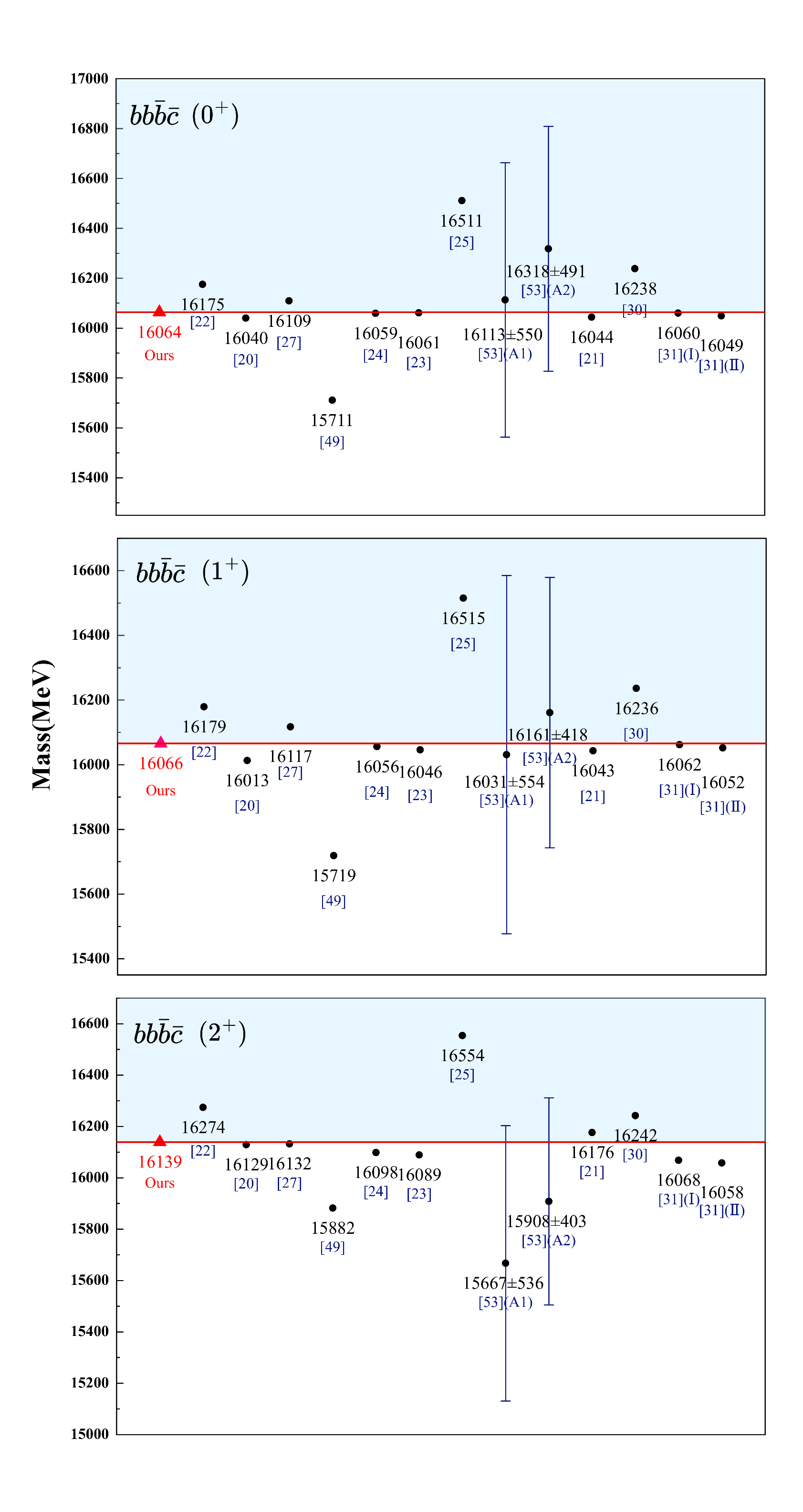}
 \caption{A comparison of the masses of the lowest $1S$-wave $bb\bar{b}\bar{c}$ states from various model predictions.}\label{bbbc comparison}
\end{figure}

For the $bb\bar{b}\bar{c}$ system, according to our quark model predictions,
there are two $J^P=0^+$ states $T_{(bb\bar{b}\bar{c})0^+}(16132)$ and $T_{(bb\bar{b}\bar{c})0^+}(16064)$,
three $J^P=1^+$ states $T_{(bb\bar{b}\bar{c})1^+}(16126)$, $T_{(bb\bar{b}\bar{c})1^+}(16119)$, and $T_{(bb\bar{b}\bar{c})1^+}(16066)$,
and one $J^P=2^+$ state $T_{(bb\bar{b}\bar{c})2^+}(16139)$. From Table~\ref{numerical results of the mass spectrum},
one can find that these predicted states should be compact states with
root-mean-square distances between any two inner quarks in the range of $(0.27, 0,40)$ fm. For comparison, our predicted masses of the lowest $1S$-wave $bb\bar{b}\bar{c}$ states, together with those of other theoretical predictions, are shown in Fig.~\ref{bbbc comparison}.
Our results are compatible with the nonrelativistic quark model predictions based on dynamic calculations
~\cite{3Deng:2020iqw,11Hu:2022zdh,12Zhang:2022qtp,35An:2022qpt} and diffusion Monte Carlo calculations~\cite{4Gordillo:2020sgc}, the diquark model predictions~\cite{5Faustov:2020qfm,Mohan:2026blk,Mutuk:2022nkw}, and the results predicted by the CGAN framework~\cite{26Malekhosseini:2025hyx}.
It should be mentioned that the results obtained with the complex scaling method~\cite{21Wu:2024hrv} are systematically $\sim450$~MeV larger than ours. Since this difference is a typical radial excitation energy, we wonder whether these resonance states obtained in~\cite{21Wu:2024hrv} may be $2S$-wave $bb\bar{b}\bar{c}$ states, and the situation is similar in other systems. On the other hand, the color-magnetic model~\cite{6Weng:2020jao} gives lower masses due to the overlooks of quark confinement potentials.


\subsubsection{$0^+$ states}

For the two $0^+$ states $T_{(bb\bar{b}\bar{c})0^+}(16132)$ and $T_{(bb\bar{b}\bar{c})0^+}(16064)$,
there is a significant mass splitting, $\Delta M\simeq 70$~MeV, which is mainly due to the spin-spin interactions.
They are mixed states between two different color configurations $6\otimes\bar{6}$ and $\bar{3}\otimes 3$.
The high mass state $T_{(bb\bar{b}\bar{c})0^+}(16132)$ is dominated by
the $\bar{3}\otimes 3$, while the low mass state $T_{(bb\bar{b}\bar{c})0^+}(16064)$ is dominated by
the $6\otimes\bar{6}$. More details can be found in Table~\ref{numerical results of the mass spectrum}. It should be mentioned that with more reliable trial wave function, the dominant components of color configurations for the low and high mass states what we obtained in the present work are different from that our previous work~\cite{ms100:2019}, except for the a notably overall mass shift.

The $T_{(bb\bar{b}\bar{c})0^+}(16132)$ and $T_{(bb\bar{b}\bar{c})0^+}(16064)$
lie about 400 MeV above the $\eta_b B_c^-$ mass threshold. Their allowed fall-apart
decay channels are $\eta_b B_c^-$ and $\Upsilon B_c^{*-}$. The fall-apart decay properties are given in Table~\ref{decay widths}.
It is seen that both $T_{(bb\bar{b}\bar{c})0^+}(16132)$ and $T_{(bb\bar{b}\bar{c})0^+}(16064)$
are predicted to be very narrow states with comparable fall-apart widths of $\sim 1$ MeV.
They may have large decay rates into the $\Upsilon B_c^*$ channel via the fall-apart decays. The partial widths
are predicted to be
\begin{eqnarray}
\Gamma [T_{(bb\bar{b}\bar{c})0^+}(16064)\to\Upsilon B_c^*]\simeq 1.1~\mathrm{MeV},\\
\Gamma [T_{(bb\bar{b}\bar{c})0^+}(16132)\to\Upsilon B_c^*]\simeq 0.44~\mathrm{MeV}.
\end{eqnarray}
For $T_{(bb\bar{b}\bar{c})0^+}(16064)$, the decay rate into the $\eta_bB_c$ channel
is also sizeable, and the partial width ratio between $\Upsilon B_c^*$ and $\eta_bB_c$ is predicted to be
\begin{eqnarray}
\mathcal{R}=\frac{\Gamma[T_{(bb\bar{b}\bar{c})0^+}(16064)\to\Upsilon B_c^*]}{\Gamma[T_{(bb\bar{b}\bar{c})0^+}(16064)\to\eta_bB_c]}\simeq 9.4.
\end{eqnarray}

\subsubsection{$1^+$ states}

Among the three $1^+$ states, the two high-lying states $T_{(bb\bar{b}\bar{c})1^+}(16126)$ and $T_{(bb\bar{b}\bar{c})1^+}(16119)$
are nearly degenerate together. There is a significant mass gap $\Delta M\simeq 50$~MeV between them and the low-lying state $T_{(bb\bar{b}\bar{c})1^+}(16066)$.
The configuration mixing in these states is slight.
As shown in Table~\ref{numerical results of the mass spectrum}, the low-lying state $T_{(bb\bar{b}\bar{c})1^+}(16066)$ is dominated by the
$6\otimes\bar{6}$ configuration, while the two high-lying states $T_{(bb\bar{b}\bar{c})1^+}(16126)$
and $T_{(bb\bar{b}\bar{c})1^+}(16119)$ are dominated by the $\bar{3}\otimes 3$ configurations $|{bb}_1^{\bar{3}}(\bar{b}\bar{c})_1^3\rangle_1^0$ and $|{bb}_1^{\bar{3}}(\bar{b}\bar{c})_0^3\rangle_1^0$, respectively.

The decay properties are given in Table~\ref{decay widths}. It is seen that both the $T_{(bb\bar{b}\bar{c})1^+}(16066)$ and $T_{(bb\bar{b}\bar{c})1^+}(16119)$ states are
narrow states with comparable widths of a few MeV.
They have significant decay rates into the $\Upsilon B_c^{-}$, $\Upsilon B_c^{*-}$, and $\eta_b B_c^{*-}$ channels.
The partial widths are predicted to be
\begin{eqnarray}
\Gamma [T_{(bb\bar{b}\bar{c})1^+}(16066)&\to &\Upsilon B_c/\Upsilon B_c^*/\eta_b B_c^{*}]\nonumber\\
&&\simeq 0.37/0.78/0.70~\mathrm{MeV},\\
\Gamma [T_{(bb\bar{b}\bar{c})1^+}(16119)&\to &\Upsilon B_c/\Upsilon B_c^*/\eta_b B_c^{*}]\nonumber\\
&&\simeq 0.66/1.34/0.96~\mathrm{MeV}.
\end{eqnarray}
The $\Upsilon B_c$ could be a promising channel for searching for these two $1^+$
$bb\bar{b}\bar{c}$ states. While for the other high-lying state $T_{(bb\bar{b}\bar{c})1^+}(16126)$, the partial widths of
the $\Upsilon B_c^{-}$, $\Upsilon B_c^{*-}$, and $\eta_b B_c^{*-}$ channels are two orders of magnitude
smaller than those of the $T_{(bb\bar{b}\bar{c})1^+}(16066)$ and $T_{(bb\bar{b}\bar{c})1^+}(16119)$.
It indicates that experimental observation of the $T_{(bb\bar{b}\bar{c})1^+}(16126)$ state
via the fall-apart decays may be challenging.

\subsubsection{$2^+$ state}

For the $2^+$ state $T_{(bb\bar{b}\bar{c})2^+}(16139)$,
as a pure $|{bb}_1^{\bar{3}}(\bar{b}\bar{c})_1^3\rangle_2^0$ state, whose mass is
very close to that of the high-lying $0^+$ and $1^+$ states,
$T_{(bb\bar{b}\bar{c})0^+}(16132)$ and $T_{(bb\bar{b}\bar{c})1^+}(16126)$.

The $\Upsilon B_c^*$ is the only allowed fall-apart decay channel
of $T_{(bb\bar{b}\bar{c})2^+}(16139)$. The partial width is predicted to be
\begin{eqnarray}
\Gamma [T_{(bb\bar{b}\bar{c})2^+}(16139)\to \Upsilon B_c^*]\simeq 0.86~\mathrm{MeV}.
\end{eqnarray}

\begin{figure}[htbp]
 \centering \epsfxsize=8.8 cm \epsfbox{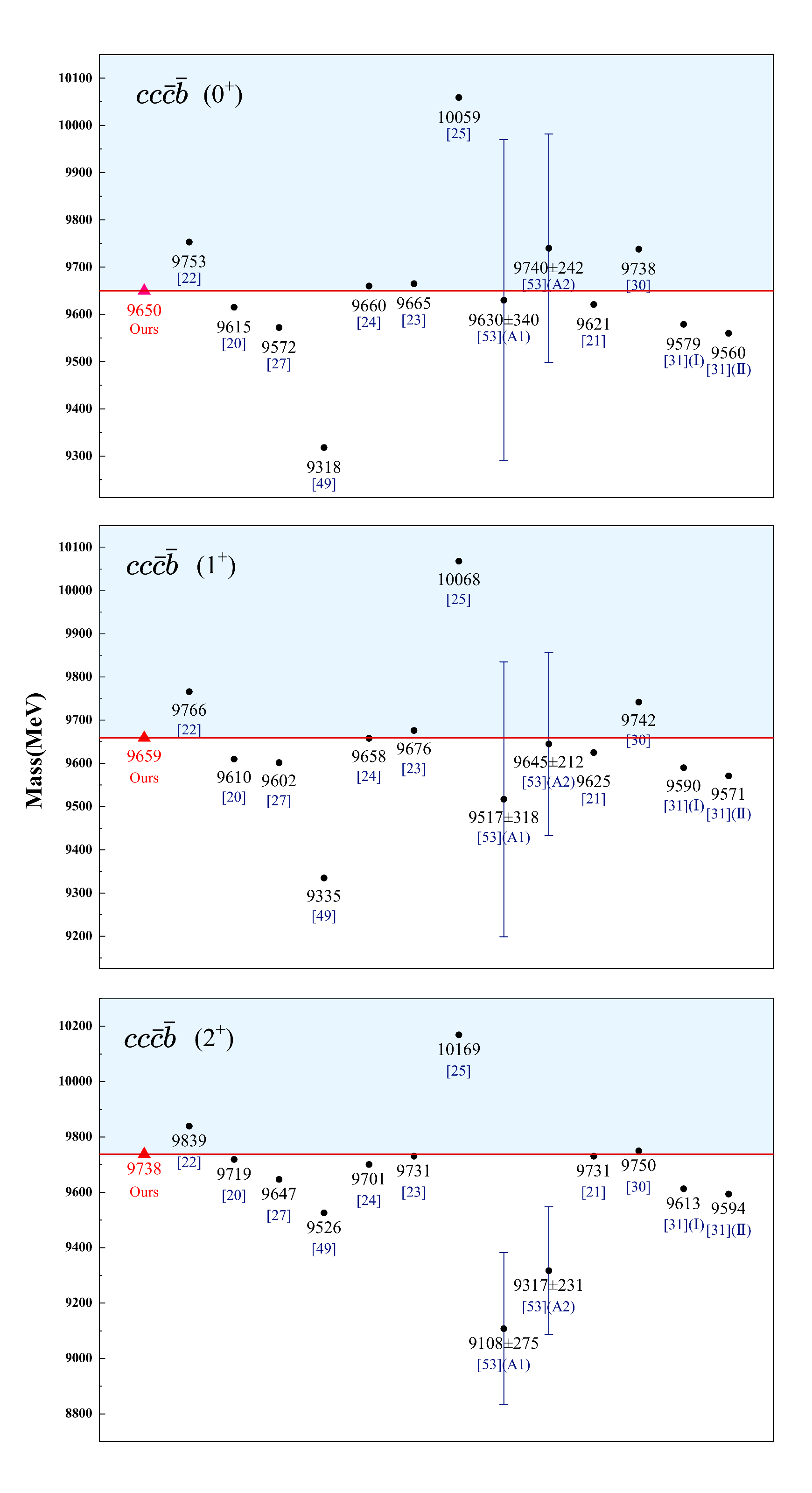}
 \caption{A comparison of the masses of the lowest $1S$-wave $cc\bar{c}\bar{b}$ states from various model predictions.}\label{cccb comparison}
\end{figure}

\subsection{$cc\bar{c}\bar{b}$}


The $cc\bar{c}\bar{b}$ system is analogous to the $bb\bar{b}\bar{c}$ system due to the same symmetry.
There are two $J^P=0^+$ states $T_{(cc\bar{c}\bar{b})0^+}(9733)$ and $T_{(cc\bar{c}\bar{b})0^+}(9650)$,
three $J^P=1^+$ states $T_{(cc\bar{c}\bar{b})1^+}(9723)$, $T_{(cc\bar{c}\bar{b})1^+}(9722)$, and $T_{(cc\bar{c}\bar{b})1^+}(9659)$,
and one $J^P=2^+$ state $T_{(cc\bar{c}\bar{b})2^+}(9738)$. From Table~\ref{numerical results of the mass spectrum}, one can
see that these states are compact states with
root-mean-square distances between any two inner quarks in the range of $(0.41, 0,51)$ fm.
For comparison, our predicted masses of the lowest $1S$-wave $cc\bar{c}\bar{b}$ states together
with those of other theoretical predictions are shown in Fig.~\ref{cccb comparison}.
Similar to the $bb\bar{b}\bar{c}$ system,
for the $cc\bar{c}\bar{b}$ system, our results are generally compatible with the nonrelativistic quark model predictions based on dynamic calculations~\cite{3Deng:2020iqw,11Hu:2022zdh,12Zhang:2022qtp,35An:2022qpt} and diffusion Monte Carlo calculations~\cite{4Gordillo:2020sgc}, the diquark model predictions~\cite{5Faustov:2020qfm,Mohan:2026blk,Mutuk:2022nkw}, and the results predicted by the CGAN framework~\cite{26Malekhosseini:2025hyx}.
It should be noted that the complex scaling method~\cite{21Wu:2024hrv} again gives higher masses for the same reason as discussed previously, while the color-magnetic model~\cite{6Weng:2020jao} shows the same lower-mass deviation as in the $bb\bar{b}\bar{c}$ system.


\subsubsection{$0^+$ states}

For the two $T_{(cc\bar{c}\bar{b})0^+}(9733)$ and $T_{(cc\bar{c}\bar{b})0^+}(9650)$,
the mass splitting is predicted to be $\Delta M\simeq 90$~MeV. The mass splitting between
$T_{(bb\bar{b}\bar{c})0^+}(16132)$ and $T_{(bb\bar{b}\bar{c})0^+}(16064)$, $\Delta M\simeq 70$~MeV,
is lightly smaller than that of the $cc\bar{c}\bar{b}$ system is due to the suppression of the heavy bottom quark.
As shown in Table~\ref{numerical results of the mass spectrum},
the $T_{(cc\bar{c}\bar{b})0^+}(9733)$ and $T_{(cc\bar{c}\bar{b})0^+}(9650)$ as mixed states, are dominated
by the $\bar{3}\otimes 3$ and $6\otimes\bar{6}$ components, respectively.
The configuration mixing for the $cc\bar{c}\bar{b}$ system is slightly stronger than that
of the $bb\bar{b}\bar{c}$ system, due to a stronger spin-spin interaction.

The decay properties are given in Table~\ref{decay widths}. One can see that both the two $0^+$ states have a narrow fall-apart decay
width of about $3$ MeV. The low-lying state $T_{(cc\bar{c}\bar{b})0^+}(9650)$ dominantly decays
into the $\eta_c B_c$ and $J/\psi B_c^*$ channels with partial decay widths of
\begin{eqnarray}
\Gamma [T_{(cc\bar{c}\bar{b})0^+}(9650)\to J/\psi B_c^*/\eta_c B_c]\simeq 0.85/1.5~\mathrm{MeV}.
\end{eqnarray}
While the high-lying $0^+$ state $T_{(cc\bar{c}\bar{b})0^+}(9733)$ has a significant decay rate
into the $J/\psi B_c^*$ channel with a partial decay width of
\begin{eqnarray}
\Gamma [T_{(cc\bar{c}\bar{b})0^+}(9733)\to J/\psi B_c^*]\simeq 2.65~\mathrm{MeV},
\end{eqnarray}
which is about a factor 3 larger than that of $T_{(cc\bar{c}\bar{b})0^+}(9650)\to J/\psi B_c^*$.
The $\eta_c B_c$ and $J/\psi B_c^*$ could be promising channels for searching for the $0^+$ $cc\bar{c}\bar{b}$ states.

\subsubsection{$1^+$ states}

The two high-lying $1^+$ states $T_{(cc\bar{c}\bar{b})1^+}(9722)$ and $T_{(cc\bar{c}\bar{b})1^+}(9723)$ are highly degenerate.
There is a significant mass gap $\Delta M\simeq 40$~MeV between them and the low-lying state
$T_{(cc\bar{c}\bar{b})1^+}(9659)$ originating from the difference of color structure.
Sizeable configuration mixing exists in these $1^+$ states.
As shown in Table~\ref{numerical results of the mass spectrum}, the low-lying state
$T_{(cc\bar{c}\bar{b})1^+}(9659)$ as a $6\otimes\bar{6}$ dominant state, also contains sizeable
$\bar{3}\otimes 3$ component. While for the two high-lying states $T_{(cc\bar{c}\bar{b})1^+}(9722)$ and $T_{(cc\bar{c}\bar{b})1^+}(9723)$, except for their dominant $\bar{3}\otimes 3$ components $|{cc}_1^{\bar{3}}(\bar{c}\bar{b})_1^3\rangle_1^0$ and $|{cc}_1^{\bar{3}}(\bar{c}\bar{b})_0^3\rangle_1^0$, they also contain a sizeable $6\otimes\bar{6}$ component.

As shown in Table~\ref{decay widths}, the two high-lying states $T_{(cc\bar{c}\bar{b})1^+}(9722)$ and $T_{(cc\bar{c}\bar{b})1^+}(9723)$
have a comparable fall-apart decay width of $\sim1$ MeV, and dominantly decay the $\eta_c B_c^*$ and $J/\psi B_c$, respectively. The partial decay
widths are predicted to be
\begin{eqnarray}
\Gamma [T_{(cc\bar{c}\bar{b})1^+}(9722)\to \eta_c B_c^*]\simeq 1.08~\mathrm{MeV},\\
\Gamma [T_{(cc\bar{c}\bar{b})1^+}(9723)\to J/\psi B_c]\simeq 0.60~\mathrm{MeV}.
\end{eqnarray}
While the low-lying state $T_{(cc\bar{c}\bar{b})1^+}(9659)$ may have sizeable decay rates into $\eta_c B_c^*$,
$J/\psi B_c$, and $J/\psi B_c^*$ channels with a comparable partial width of $\sim 0.1-0.2$ MeV.
The $J/\psi B_c$ could be a promising channel for searching for the $1^+$ states
$T_{(cc\bar{c}\bar{b})1^+}(9723)$ and $T_{(cc\bar{c}\bar{b})1^+}(9659)$.

\subsubsection{$2^+$ state}

For the $2^+$ state $T_{(cc\bar{c}\bar{b})2^+}(9738)$,
as a pure $|{cc}_1^{\bar{3}}(\bar{c}\bar{b})_1^3\rangle_2^0$ state, the mass is
very close to that of the high-lying $0^+$ and $1^+$ states, $T_{(cc\bar{c}\bar{b})0^+}(9733)$ and $T_{(cc\bar{c}\bar{b})1^+}(9723)$.

The $J/\psi B_c^*$ is the only allowed fall-apart decay channel
in all of $T_{(cc\bar{c}\bar{b})2^+}(9738)$. The
partial width is predicted to be
\begin{eqnarray}
\Gamma [T_{(cc\bar{c}\bar{b})2^+}(9738)\to J/\psi B_c^*]\simeq 0.18~\mathrm{MeV},
\end{eqnarray}
which is comparable with that of $T_{(cc\bar{c}\bar{b})1^+}(9659,9723)\to J/\psi B_c^*$,
however, is about an order of magnitude
smaller than that of $T_{(cc\bar{c}\bar{b})0^+}(9650,9733)\to J/\psi B_c^*$.
Thus, compared to these $0^+$ states, the $2^+$ state $T_{(cc\bar{c}\bar{b})2^+}(9738)$
may be more difficult to discover in the $J/\psi B_c^*$ channel.

\begin{figure}[htbp]
 \centering \epsfxsize=9 cm \epsfbox{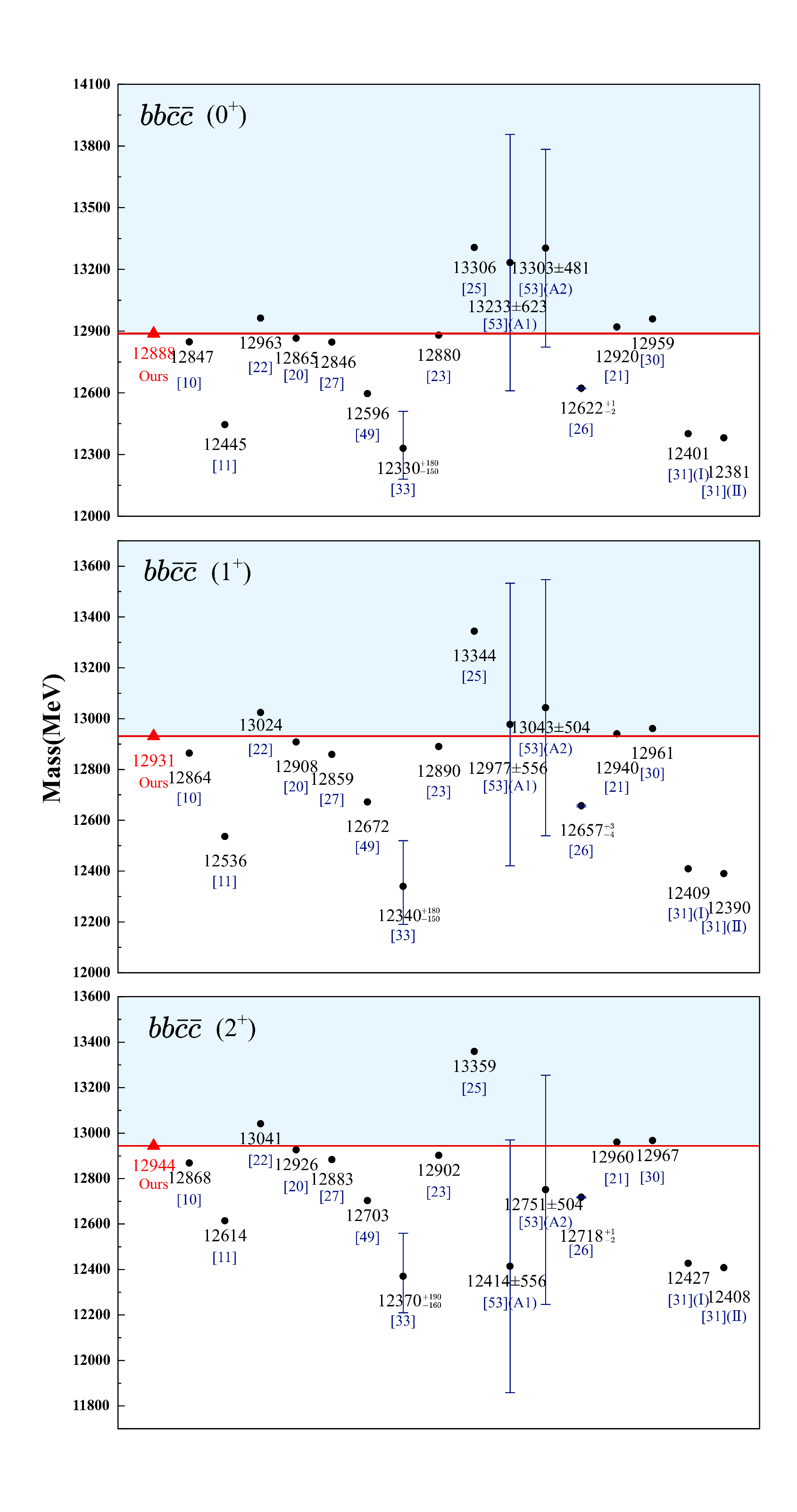}
 \caption{A comparison of the masses of the lowest $1S$-wave $bb\bar{c}\bar{c}$ states from various model predictions.}\label{bbcc comparison}
\end{figure}

\begin{figure}[htbp]
 \centering \epsfxsize=9 cm \epsfbox{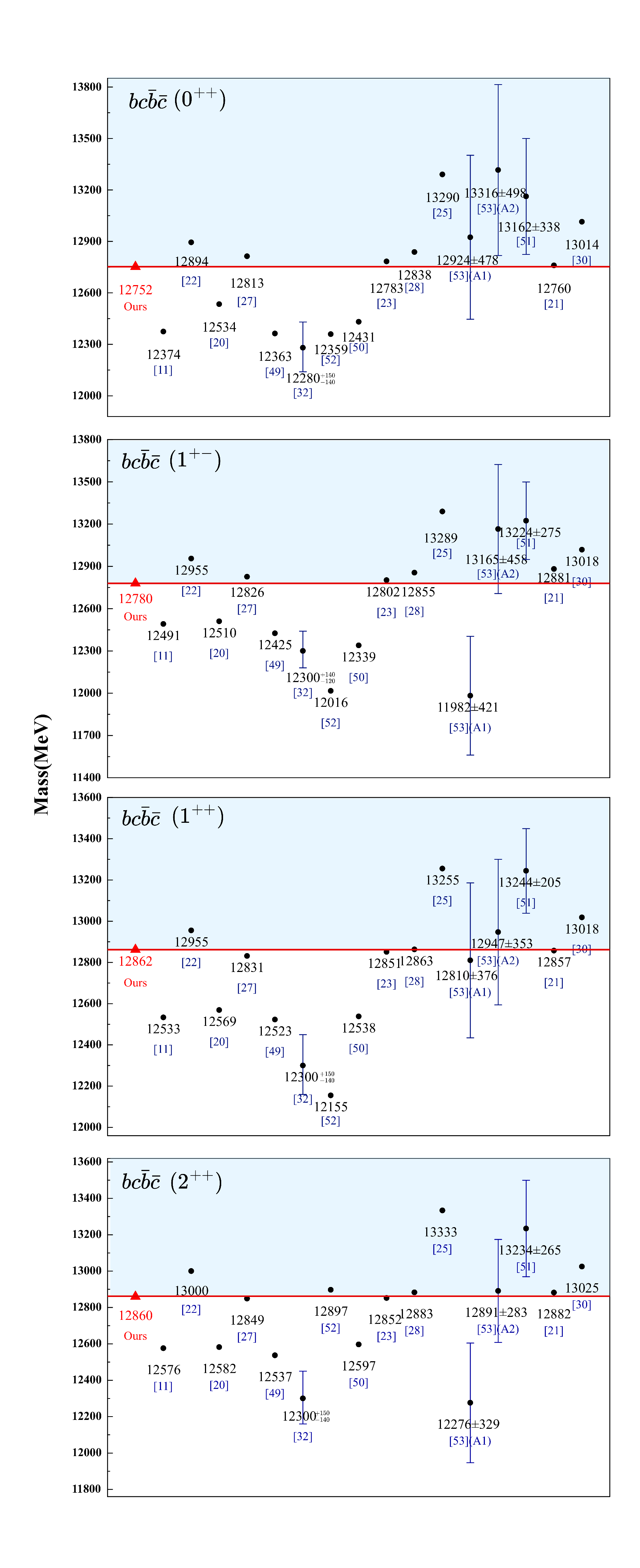}
\caption{A comparison of the masses of the lowest $1S$-wave $bc\bar{b}\bar{c}$ states from various model predictions.}\label{bcbc comparison}
\end{figure}

\subsection{$bb\bar{c}\bar{c}$}

For the $bb\bar{c}\bar{c}$ system, according to our quark model predictions,
there are two $J^P=0^+$ states $T_{(bb\bar{c}\bar{c})0^+}(12942)$ and $T_{(bb\bar{c}\bar{c})0^+}(12888)$,
one $J^P=1^+$ state $T_{(bb\bar{c}\bar{c})1^+}(12931)$,
and one $J^P=2^+$ state $T_{(bb\bar{c}\bar{c})2^+}(12944)$.
From Table~\ref{numerical results of the mass spectrum}, it is seen that these four states highly overlap within a
very small mass region $(12.88,12.95)$ GeV. They should be compact states with
root-mean-square distances between any two inner quarks in the range of $(0.28, 0,48)$ fm.
For comparison, our predicted masses of the lowest $1S$-wave $bb\bar{c}\bar{c}$ states, together with those of other theoretical predictions, are shown in Fig.~\ref{bbcc comparison}. Our results are generally compatible with the nonrelativistic quark model predictions based on dynamic calculations~\cite{1Wang:2019rdo,3Deng:2020iqw,12Zhang:2022qtp,35An:2022qpt} and diffusion Monte Carlo calculations~\cite{4Gordillo:2020sgc},
the diquark model predictions~\cite{5Faustov:2020qfm,Mohan:2026blk}, and the results predicted by the CGAN framework~\cite{26Malekhosseini:2025hyx}.
It should be noted that the complex scaling method~\cite{21Wu:2024hrv} again gives higher masses for the same reason as discussed above. Several approaches give lower masses: the diquark models~\cite{2Bedolla:2019zwg,Mutuk:2022nkw}, which assume a strong diquark correlation, and the QCD sum rule calculation~\cite{9Wang:2021taf}, which employs diquark-type interpolating currents, as well as the color-magnetic model~\cite{6Weng:2020jao} which neglects dynamical confinement.


\subsubsection{$0^+$ states}

For the two $0^+$ states, $T_{(bb\bar{c}\bar{c})0^+}(12942)$ and $T_{(bb\bar{c}\bar{c})0^+}(12888)$,
there is a significant mass splitting of $\Delta M\sim50$~MeV.
As shown in Table~\ref{numerical results of the mass spectrum}, they are mixed states between
two different color configurations. The high and low mass states are dominated by
the $|\{bb\}_1^{\bar{3}}(\bar{c}\bar{c})_1^3\rangle_0^0$ and $|\{bb\}_0^6(\bar{c}\bar{c})_0^{\bar{6}}\rangle_0^0$ components, respectively.
The mass of $|\{bb\}_0^6(\bar{c}\bar{c})_0^{\bar{6}}\rangle_0^0$ is smaller than
that of $|\{bb\}_1^{\bar{3}}(\bar{c}\bar{c})_1^3\rangle_0^0$, which is consistent with the prediction of model I in Ref.~\cite{1Wang:2019rdo}.

The $T_{(bb\bar{c}\bar{c})0^+}(12942)$ and $T_{(bb\bar{c}\bar{c})0^+}(12888)$ lie about $300$ MeV
above the mass threshold of $B_c^{*-}B_c^{*-}$. Their allowed fall-apart decay channels are $B_c^-B_c^-$ and $B_c^{*-}B_c^{*-}$.
As shown in Table~\ref{decay widths}, the $T_{(bb\bar{c}\bar{c})0^+}(12942)$ and $T_{(bb\bar{c}\bar{c})0^+}(12888)$ may be narrow states
with widths of $\sim2.0$~MeV and $\sim0.5$~MeV, respectively.
The $T_{(bb\bar{c}\bar{c})0^+}(12942)$ has comparable decay rates into both $B_c^*B_c^*$ and $B_cB_c$ channels.
The partial widths are predicted to be
\begin{eqnarray}
\Gamma [T_{(bb\bar{c}\bar{c})0^+}(12942)\to B_cB_c/B_c^*B_c^*]\simeq 1.12/0.87~\mathrm{MeV},
\end{eqnarray}
The low mass $0^+$ state $T_{(bb\bar{c}\bar{c})0^+}(12888)$
dominantly decays into $B_cB_c$ channel with a partial width of
\begin{eqnarray}
\Gamma [T_{(bb\bar{c}\bar{c})0^+}(12888)\to B_cB_c]\simeq 0.37~\mathrm{MeV},
\end{eqnarray}
while the decay rate into the $B_c^*B_c^*$ channel is sizeable.
The partial width ratio between these two channels is predicted to be
\begin{eqnarray}
\mathcal{R}=\frac{\Gamma[T_{(bb\bar{c}\bar{c})0^+}(12888)\to B_c^*B_c^*]}
{\Gamma[T_{(bb\bar{c}\bar{c})0^+}(12888)\to B_cB_c]}\simeq 0.4.
\end{eqnarray}
The $B_cB_c$ could be a promising channel for searching for the $0^+$ $bb\bar{c}\bar{c}$ states in experiments.

\subsubsection{$1^+$ and $2^+$ states}

The $T_{(bb\bar{c}\bar{c})2^+}(12944)$, as the highest mass state in
the $bb\bar{c}\bar{c}$ system, only about $13$ MeV lies above the $1^+$
state $T_{(bb\bar{c}\bar{c})1^+}(12931)$, and is also nearly degenerate
with the high-lying $0^+$ state $T_{(bb\bar{c}\bar{c})0^+}(12942)$, due to the similar color-spin structures.

For the $T_{(bb\bar{c}\bar{c})1^+}(12931)$ and $T_{(bb\bar{c}\bar{c})2^+}(12944)$,
the allowed fall-apart decay channels are $B_cB_c^*$ and $B_c^*B_c^*$, respectively.
The partial widths are predicted to be
\begin{eqnarray}
\Gamma [T_{(bb\bar{c}\bar{c})1^+}(12931)\to B_cB_c^*]\simeq 0.10~\mathrm{MeV},\\
\Gamma [T_{(bb\bar{c}\bar{c})2^+}(12944)\to B_c^*B_c^*]\simeq 1.52~\mathrm{MeV}.
\end{eqnarray}

\subsection{$bc\bar{b}\bar{c}$}

For the $bc\bar{b}\bar{c}$ system, due to no constraints from the Pauli principle,
there are more states than the other systems.
According to our quark model calculations, we obtain four $J^P=0^{++}$ states $T_{(bc\bar{b}\bar{c})0^{++}}(12985/12936/12853/12752)$,
four $J^P=1^{+-}$ states $T_{(bc\bar{b}\bar{c})1^{+-}}(12987/12970/12826/12780)$,
two $J^P=1^{++}$ states $T_{(bc\bar{b}\bar{c})1^{++}}(12945/12862)$,
and two $J^P=2^{++}$ states $T_{(bc\bar{b}\bar{c})2^{++}}(12981/12860)$.
These twelve states scatter in a relatively large mass region $(12.75,12.99)$ GeV.
As shown in Table~\ref{numerical results of the mass spectrum}, they should be compact states with root-mean-square distances between any two inner quarks in the range of $(0.29, 0,47)$ fm.
For comparison, our predicted masses of the lowest $1S$-wave $bc\bar{b}\bar{c}$ states together
with those of other theoretical predictions are shown in Fig.~\ref{bcbc comparison}. Our results are generally compatible with the nonrelativistic quark model predictions based on dynamic calculations~\cite{3Deng:2020iqw,12Zhang:2022qtp,35An:2022qpt},
the diquark model predictions~\cite{5Faustov:2020qfm,13Faustov:2022mvs}, and the results predicted by the CGAN framework~\cite{26Malekhosseini:2025hyx}.
It should be noted that the complex scaling method~\cite{21Wu:2024hrv} again gives higher masses for the same reason as discussed previously, while the Bethe-Salpeter equation approach~\cite{30Wang:2025apq} gives higher masses due
to introducing a different interaction kernel. Several approaches give lower masses: the diquark model~\cite{2Bedolla:2019zwg}, which assumes a strong diquark correlation; the QCD sum rule calculation~\cite{7Yang:2021zrc}, which employs diquark-type interpolating currents; the color-magnetic models~\cite{6Weng:2020jao,10Zhuang:2021pci} which neglect dynamical confinement; and the bosonic algebraic approach~\cite{8Majarshin:2021hex} which adopts a different parametrization of the dynamics.

\subsubsection{$0^{++}$ states}

For the four $J^P=0^{++}$ states, there are strong configuration mixings
between the $6\otimes\bar{6}$ and $\bar{3}\otimes 3$ configurations.
From Table~\ref{numerical results of the mass spectrum}, one can find that the dominant color component of the two high-lying states $T_{(bc\bar{b}\bar{c})0^{++}}(12985)$ and
$T_{(bc\bar{b}\bar{c})0^{++}}(12936)$ is $\bar{3}\otimes 3$. While for the two low-lying states
$T_{(bc\bar{b}\bar{c})0^{++}}(12853)$ and $T_{(bc\bar{b}\bar{c})0^{++}}(12752)$,
the dominant color component is $6\otimes\bar{6}$. There is a significant mass interval, $\Delta M\sim 50-100$ MeV, between
any two adjacent states.

As shown in Table~\ref{decay widths}, the two low-lying $0^{++}$ states $T_{(bc\bar{b}\bar{c})0^{++}}(12752)$ and $T_{(bc\bar{b}\bar{c})0^{++}}(12853)$
have narrow fall-apart widths of $\sim 3$, and $\sim 1$ MeV, respectively.
The partial widths of their main decay channels are predicted to be
\begin{eqnarray}
\Gamma [T_{(bc\bar{b}\bar{c})0^{++}}(12752)&\to& \eta_b\eta_c/B_c^+B_c^-/B_c^{*+}B_c^{*-}]\nonumber\\
&&~~~~\simeq 0.37/0.51/2.4~\mathrm{MeV}.\\
\Gamma [T_{(bc\bar{b}\bar{c})0^{++}}(12853)&\to& \eta_b\eta_c/\Upsilon J/\psi/B_c^{*+}B_c^{*-}]\nonumber\\
&&~~~~\simeq 0.10/0.50/0.39~\mathrm{MeV}.
\end{eqnarray}
The two high-lying $0^{++}$ states $T_{(bc\bar{b}\bar{c})0^{++}}(12985)$ and $T_{(bc\bar{b}\bar{c})0^{++}}(12936)$
have comparable fall-apart widths of $\sim 3$ MeV.
The $T_{(bc\bar{b}\bar{c})0^{++}}(12985)$ mainly decays into $\Upsilon J/\psi$ and $B_c^+B_c^-$ channels
with partial widths of
\begin{eqnarray}
\Gamma [T_{(bc\bar{b}\bar{c})0^{++}}(12985)\to \Upsilon J/\psi/B_c^{+}B_c^{-}]\simeq 2.0/0.60~\mathrm{MeV}.
\end{eqnarray}
While the highest state $T_{(bc\bar{b}\bar{c})0^{++}}(12936)$ mainly decays into $\eta_b\eta_c$ and $B_c^{*+}B_c^{*-}$ channels
with partial widths of
\begin{eqnarray}
\Gamma [T_{(bc\bar{b}\bar{c})0^{++}}(12936)\to \eta_b\eta_c/B_c^{*+}B_c^{*-}]\simeq 1.6/1.7~\mathrm{MeV}.
\end{eqnarray}
The $\eta_b\eta_c$, $\Upsilon J/\psi$, and $B_c^+B_c^-$ could be promising channels for
searching for these $0^{++}$ $bc\bar{b}\bar{c}$ states in experiments.

\subsubsection{$1^{+-}$ states}

For the four $J^P=1^{+-}$ states, there are also strong configuration mixings
between the $6\otimes\bar{6}$ and $\bar{3}\otimes 3$ configurations.
The dominant color component of the two high-lying states $T_{(bc\bar{b}\bar{c})1^{+-}}(12987)$ and
$T_{(bc\bar{b}\bar{c})1^{+-}}(12970)$ is $\bar{3}\otimes 3$. While for the two low-lying states
$T_{(bc\bar{b}\bar{c})1^{+-}}(12826)$ and $T_{(bc\bar{b}\bar{c})1^{+-}}(12780)$,
the dominant color component is $6\otimes\bar{6}$. More details can be found in Table~\ref{numerical results of the mass spectrum}.

In these $1^{+-}$ states, as shown in Table~\ref{decay widths}, the lowest state $T_{(bc\bar{b}\bar{c})1^{+-}}(12780)$ has a
relatively broad fall-apart width of $\sim 0.4$ MeV. It may have sizeable decay rates into
the $B_cB_c^*=B_c^+B_c^{*-}+B_c^{*+}B_c^{-}$ and $B_c^{*+}B_c^{*-}$ channels with comparable
partial widths
\begin{eqnarray}
\Gamma [T_{(bc\bar{b}\bar{c})1^{+-}}(12780)\to B_cB_c^*/B_c^{*+}B_c^{*-}]\simeq 0.21/0.14~\mathrm{MeV}.
\end{eqnarray}
For the other $1^{+-}$ states, the fall-apart decay widths are predicted to be
$\sim100$ keV. These states may be difficult to observe in their fall-apart decay channels.

\subsubsection{$1^{++}$ and $2^{++}$ states}

From Table~\ref{numerical results of the mass spectrum}, one can find that there is a slight mixing between the $6\otimes\bar{6}$ and $\bar{3}\otimes 3$ configurations in the $1^{++}$ and $2^{++}$ states. The low-mass state $T_{(bc\bar{b}\bar{c})1^{++}}(12862)$ and the high-mass
state $T_{(bc\bar{b}\bar{c})1^{++}}(12945)$ are governed by the $6\otimes\bar{6}$ and $\bar{3}\otimes 3$ components, respectively.
However, for the $2^{++}$ sector, the case is reversed, the low-mass state $T_{(bc\bar{b}\bar{c})2^{++}}(12860)$ and the high-mass
state $T_{(bc\bar{b}\bar{c})2^{++}}(12981)$ are governed by the $\bar{3}\otimes 3$ and $6\otimes\bar{6}$ components, respectively.
The mass splitting between the two states with the same spin-parity numbers is significant, the value can reach up to $\Delta M\sim 100$ MeV.
It should be mentioned that the $T_{(bc\bar{b}\bar{c})1^{++}}(12862)$ and $T_{(bc\bar{b}\bar{c})2^{++}}(12860)$
are highly degenerate with each other due to the similar spin-color structures.

As shown in Table~\ref{decay widths}, the low-mass $2^{++}$ state $T_{(bc\bar{b}\bar{c})2^{++}}(12860)$ has a narrow fall-apart
widths of $\sim 0.4$ MeV. It has significant decay rates into both the
$\Upsilon J/\psi$ and $B_c^{*+}B_c^{*-}$ channels with partial width of
\begin{eqnarray}
\Gamma [T_{(bc\bar{b}\bar{c})2^{++}}(12860)\to \Upsilon J/\psi/B_c^{*+}B_c^{*-}]\simeq 0.16/0.27~\mathrm{MeV}.
\end{eqnarray}
This state may have potentials to be observed in future experiments.
For the two $1^{++}$ states and the high-mass $2^{++}$ state,
from Table~\ref{decay widths}, it is found that their fall-apart decay channels
are nearly forbidden. Thus, the possibility of establishing these states via the
fall-apart decay processes may be very small.

\begin{table}[hptb]
\centering
\caption{A short list of the most promising tetraquark states with their quantum numbers, mass regions, and optimal decay channels.}
\label{tab:promising_states}
\begin{tabular}{ccccc}
\hline\hline
System & $J^{P(C)}$ & Mass (MeV) & Optimal channel(s) & Width (MeV) \\
\hline
\multirow{5}{*}{$bb\bar{b}\bar{c}$}
 & $0^+$ & 16064 & $\Upsilon B_c^*$ & 1.13 \\
 & $0^+$ & 16132 & $\Upsilon B_c^*$ & 0.44 \\
 & $1^+$ & 16066 & $\Upsilon B_c$ & 0.37 \\
 & $1^+$ & 16119 & $\Upsilon B_c$ & 0.66 \\
 & $2^+$ & 16139 & $\Upsilon B_c^*$ & 0.86 \\
\hline
\multirow{4}{*}{$cc\bar{c}\bar{b}$}
 & $0^+$ & 9650 & $\eta_c B_c$ / $J/\psi B_c^*$ & 1.50 / 0.85 \\
 & $0^+$ & 9733 & $J/\psi B_c^*$ & 2.65 \\
 & $1^+$ & 9659 & $J/\psi B_c$ & 0.20 \\
 & $1^+$ & 9723 & $J/\psi B_c$ & 0.60 \\
\hline
\multirow{3}{*}{$bb\bar{c}\bar{c}$}
 & $0^+$ & 12888 & $B_c B_c$ & 0.37 \\
 & $0^+$ & 12942 & $B_c B_c$ & 1.12 \\
 & $2^+$ & 12944 & $B_c^* B_c^*$ & 1.52 \\
\hline
\multirow{6}{*}{$bc\bar{b}\bar{c}$}
 & $0^{++}$ & 12752 & $\eta_b\eta_c$ / $B_c^+ B_c^-$ & 0.37 / 0.51 \\
 & $0^{++}$ & 12853 & $\Upsilon J/\psi$ & 0.50 \\
 & $0^{++}$ & 12936 & $\eta_b\eta_c$  & 1.58 \\
 & $0^{++}$ & 12985 & $\Upsilon J/\psi$ / $B_c^+ B_c^-$ & 2.01 / 0.60 \\
 & $1^{+-}$ & 12780 & $B_c B_c^*$ / $B_c^{*+} B_c^{*-}$ & 0.21 / 0.14 \\
 & $2^{++}$ & 12860 & $\Upsilon J/\psi$ / $B_c^{*+} B_c^{*-}$ & 0.16 / 0.27 \\
\hline\hline
\end{tabular}
\end{table}

\section{SUMMARY}\label{summary}

In this work, we carry out a precise calculation of the mass spectrum
of the tetraquarks $bb\bar{b}\bar{c}$, $cc\bar{c}\bar{b}$,
$bb\bar{c}\bar{c}$, and $bc\bar{b}\bar{c}$ with a nonrelativistic potential model based
on the reliable ECG numerical method.
A complete mass spectrum for the $1S$ states is obtained.
The masses of the $1S$-wave states for the $bb\bar{b}\bar{c}$, $cc\bar{c}\bar{b}$, $bb\bar{c}\bar{c}$,
and $bc\bar{b}\bar{c}$ systems are predicted to be in the ranges $\sim(16.06,16.14)$, $\sim(9.65,9.74)$, $\sim(12.89,12.94)$, and $\sim(12.75,12.99)$~GeV, respectively.
All states are compact structures and lie significantly above their dissociation two ground meson threshold.
Compared to our previous rough predictions, it is found that the masses of all the states predicted
with the reliable ECG method are shifted downward by around $30-100$ MeV, and the mass
splittings are also notably modified.

Moreover, by using the obtained masses and wave functions of the $1S$-wave states
for the $bb\bar{b}\bar{c}$, $cc\bar{c}\bar{b}$, $bb\bar{c}\bar{c}$, and $bc\bar{b}\bar{c}$ systems,
we further evaluate the fall-apart decay properties within a quark exchange model.
The all-heavy tetraquarks are likely to be narrow states, their fall-apart widths are
predicted to range from a few tenths to several MeV. The partial widths of
the fall-apart decay channels for each $1S$-wave state are given.
Some $1S$ states for the $bb\bar{b}\bar{c}$, $cc\bar{c}\bar{b}$, $bb\bar{c}\bar{c}$, and $bc\bar{b}\bar{c}$ could be search for in their optimal fall-apart decay channels.
Based on the calculated decay widths and the feasibility of final-state reconstruction,
we list the most promising states and their optimal channels in Table~\ref{tab:promising_states}.

Finally, it should be mentioned that the states obtained in the present quark model framework (often called bare states) are not strictly physical states since the coupled-channel effects from the decay channels (two hadron continuum) are neglected as an approximation.
Fortunately, the narrow fall-apart decay widths of the quark model states indicate that the mass shift connections
from the continuum couplings should not be significant. On the other hand, the mixing with the meson-meson states
is also suppressed since the obtained bare compact states lie far above the meson-meson thresholds.
Thus, the predicted masses and widths for the bare compact states might be comparable to those of physical resonances
if they are observed in the invariant mass spectrum of final states by experiments.

\section*{Acknowledgement}
This work is supported by the Basic Research Project for Young Students of the Natural Science Foundation of Hunan Province (Grant No. 2024JJ10038),
National Students' Platform for Innovation and Entrepreneurship Training Program(S202410542033),
and the National Natural Science Foundation of China (Grant Nos. 12105203, 12235018, and 12175065).

\end{document}